\documentclass[10pt,twocolumn]{article}
\usepackage[margin=0.72in,top=0.8in,bottom=0.9in]{geometry}
\usepackage{amsmath,amssymb,amsthm,bm,mathtools}
\usepackage{graphicx}
\usepackage{booktabs}
\usepackage{microtype}
\usepackage[numbers,sort&compress]{natbib}
\usepackage[colorlinks,linkcolor=blue!60!black,citecolor=blue!60!black,urlcolor=blue!60!black]{hyperref}
\usepackage{xcolor}
\hypersetup{pdftitle={Local gluing classification and bulk-locked clock rings for Z(N) models on a Klein bottle},pdfauthor={Junhao (Eigen) Wang},pdfsubject={Non-orientable lattice models and qudit stabilizer codes}}
\makeatletter % run-in paragraph headings with less space above (article default 3.25ex)
\renewcommand\paragraph{\@startsection{paragraph}{4}{\z@}{1.6ex \@plus.6ex \@minus.2ex}{-1em}{\normalfont\normalsize\bfseries}}
\makeatother
\newtheorem{theorem}{Theorem}
\newtheorem{proposition}{Proposition}
\newtheorem{lemma}{Lemma}
\newtheorem{corollary}{Corollary}
\theoremstyle{definition}
\newtheorem{definition}{Definition}
\newtheorem{example}{Example}
\newtheorem{remark}{Remark}
\newcommand{\Z}{\mathbb Z}
\newcommand{\GSD}{\mathrm{GSD}}
\newcommand{\Pd}{P_{\mathrm d}}
\newcommand{\Pe}{P_{\mathrm e}}
\newcommand{\coker}{\operatorname{coker}}
\newcommand{\im}{\operatorname{im}}
\newcommand{\Op}{\mathcal O}
\newcommand{\cS}{\mathcal S}
\newcommand{\cL}{\mathcal L}
\newcommand{\cR}{\mathcal R}
\newcommand{\cW}{\mathcal W}
\newcommand{\cD}{\mathcal D}
\newcommand{\cF}{\mathcal F}
\newcommand{\cG}{\mathcal G}
\newcommand{\sM}{\mathsf M}
\newcommand{\Had}{\mathsf H}

\title{\bfseries\Large Local gluing classification and bulk-locked clock rings\\ for $\mathbb Z_N$ models on a Klein bottle}
\author{Junhao (Eigen) Wang\\[2pt] \small Department of Mathematics, New York University, New York, NY, USA\\ \small \texttt{jw10111@nyu.edu}}
\date{\small November 13, 2026}

\ifdefined\XeTeXversion
\DeclareFontShape{TU}{lmr}{bx}{n}{<->"[lmroman10-bold]:mapping=tex-text;"}{}
\DeclareFontShape{TU}{lmr}{m}{it}{<->"[lmroman10-italic]:mapping=tex-text;"}{}
\DeclareFontShape{OML}{cmm}{m}{it}{<-6>cmmi5<6-7>cmmi6<7-8>cmmi7<8-9>cmmi8<9->cmmi10}{}
\DeclareFontShape{OT1}{cmr}{m}{n}{<-6>cmr5<6-7>cmr6<7-8>cmr7<8-9>cmr8<9->cmr10}{}
\DeclareFontFamily{U}{msb}{}
\DeclareFontFamily{U}{msa}{}
\DeclareFontShape{U}{msb}{m}{n}{<->msbm10}{}
\DeclareFontShape{U}{msa}{m}{n}{<->msam10}{}
\DeclareFontShape{OMX}{cmex}{m}{n}{<->cmex10}{}
\DeclareFontFamily{U}{euf}{}
\DeclareFontShape{U}{euf}{m}{n}{<->eufm10}{}
\DeclareFontFamily{TU}{lmtt}{}
\DeclareFontShape{TU}{lmtt}{m}{n}{<->"[lmmono10-regular]"}{}
\fi

\begin{document}
\maketitle

\begin{abstract}
\noindent
We classify local reflection gluings of the $\mathbb Z_N$ plaquette model on lattice Klein bottles and solve the obstructed literal seam. Within the monomial far-corner ansatz, all commuting gluings form a family labelled by a unit $u$ of $\mathbb Z_N$. They exist for every $N$ at even seam length and, at odd length, exactly when $-1$ is a square modulo $N$. We determine the anyon transport, including exchange seams with $P^2=C$ for charge conjugation $C$, and the relation phases that distinguish an unfrustrated code from a commuting but frustrated Hamiltonian. For the literal gluing, the $2L_x$ noncommuting plaquettes form self-dual clock rings whose sectors are locked by the bulk. This gives an exact reduction of the spectrum to clock-ring spectra, closed-form ring ground energies for $N=2,4$, and a closed-form ground-state degeneracy. We further determine twisted-coefficient homology on the stated surface domains and compare it with logical modules on registered lattice grids; composite dimensions need not give squares of cyclic groups. Dressed-isometry gates and monodromy are reported with their finite-grid scope. Finally, the qudit Gottesman--Knill theorem bounds the computational interpretation: Clifford deformations and Weyl measurements with stabilizer inputs and no external magic states cannot supply a non-stabilizer resource. Charge-conjugation measurements that do supply such states lie outside this operation model. The new results concern explicit seam classification, phase-sensitive consistency and bulk-locked spectra, not a new general no-go theorem or a universal computing architecture.
\end{abstract}

\section{Introduction}\label{sec:intro}

A reflection gluing can preserve a lattice check algebra, frustrate its vacuum sector through a relation phase, or produce a noncommuting seam Hamiltonian. These possibilities are distinct even when the bulk topological order is unchanged. We study them explicitly for the $\Z_N$ plaquette model on a Klein bottle. Our main results are a classification of commuting local gluings within a stated ansatz and an exact reduction of the obstructed literal seam to bulk-locked clock rings.

In Ref.~\cite{Wang2026paper3} we studied the $\Z_2$ toric code on a lattice Klein bottle. In the edge register the reflection gluing preserves the checks; in the site register (Wen's plaquette model~\cite{Wen2003plaquette}) a Hadamard dressing restores commutation, while the literal gluing produces locked Majorana rings. The extension to $\Z_N$ adds structures absent at $N=2$: a unit-labelled $u$-family, an arithmetic obstruction at odd circumference, order-four exchange transport with $P^2=C$, and logical modules with composite-$N$ torsion. Here $C$ is charge conjugation on anyon labels. None of these label relations alone specifies a measurement or a non-Clifford resource.

\paragraph{Contributions and results map.} Theorems~\ref{thm:class} and~\ref{thm:deg} classify all commuting gluings within the local monomial far-corner ansatz, determine their seam actions and code dimensions, and track the relation phases needed for an unfrustrated code. Theorems~\ref{thm:anatomy} and~\ref{thm:lock} reduce the noncommuting literal seam to self-dual $\Z_N$ clock rings with bulk-constrained sectors. These are the two central microscopic results. Theorem~\ref{thm:homology}, on its stated domain, and the separate projective-plane calculation in Example~\ref{ex:rp2} describe twisted homology. Propositions~\ref{prop:L}, \ref{prop:gates} and~\ref{prop:mono} compare logical modules, dressed isometries and monodromy with their stated proof, argument and grid support. Finally, Corollary~\ref{cor:nogo} and Remark~\ref{rem:opboundary} delimit the computation allowed by the operation model below; Proposition~\ref{prop:wigner} identifies the non-stabilizer measurements in the Wigner examples.

\paragraph{Relation to known results.} Non-orientable qudit homological codes and parity-invariant-anyon counts are known~\cite{Lei2023qudit,ChanTeoRyu2016}, as are reflection-enriched phases~\cite{BBCJW2020} and cross-cap Clifford gates at $N=2$~\cite{Kobayashi2024crosscap}. Recent quotient-code methods connect twisted compactification to logical operators and gates~\cite{Hsin2026quotient}. We do not claim these general principles, the mere possibility of order-four reflection transport, or stabilizer closure as new. Our contribution is their resolution in the specified plaquette realization: the complete local $u$-gluing classification, its phase-sensitive consistency conditions, and the locked clock-ring spectrum of the literal obstruction. The composite-$N$ and gate results retain the evidence levels of Table~\ref{tab:scope}. Section~\ref{sec:lit} compares the constructions and assumptions in detail. The $N=2$ results of Ref.~\cite{Wang2026paper3} are re-derived rather than re-claimed.

\paragraph{Operation model and computational boundary.} For the computational discussion we specify $\Op$: static codes generated by $\Z_N$ Weyl stabilizers; code deformations implemented by Clifford unitaries and Weyl measurements; classical feed-forward; stabilizer inputs and no external non-stabilizer states. Lattice isometries with single-site Clifford dressings and cross-cap moves belong to $\Op$ only when implemented by these operations. The qudit Gottesman--Knill theorem then excludes production of a non-stabilizer state, independently of the seam classification. In proposals using $\Z_3$ topological order, additional resources enter through a symmetry-constrained charge measurement~\cite{CongChengWang2017} or coherent interferometry with charge-conjugation defects~\cite{ShiHanRaghuVishwanath2026}. Extra symmetry and fractionalization data, cross-cap-state preparation, coherent superpositions of mobile-defect paths, dynamical non-Pauli measurements and non-Clifford pumps are not supplied by our static gluing and lie outside $\Op$.

\paragraph{Scope.} The classification is restricted to local monomial far-corner gluings of the fixed bulk plaquette model. We construct lattice codes only on the alternating Klein bottle; elsewhere we compute homology without asserting a lattice realization. In the exchange branch the double-winding external monodromy is $P^{\pm2}=C$, but a single passage is $P$, and at $N=2$ the exchange $e\leftrightarrow m$ is not $C=1$. Fusion counting is a consistency check, not an implementation of a defect theory. The no-go applies to $\Op$, not to arbitrary physical manipulations of cross-caps. No protected non-Clifford implementation, noise threshold or universal architecture is established.

\paragraph{Method and evidence.} The work was done in five stages: E-0, an engine calibrated against the $N=2$ results of Ref.~\cite{Wang2026paper3} (no pre-registration), and E-1 to E-4, whose predictions and grids were frozen before the stage's runs; results known beforehand are listed in each pre-registration and not counted as predictions. Each final run was sealed by sha256 after development or rehearsal runs (Appendix~\ref{app:repro}). Table~\ref{tab:scope} labels each result by its support---\emph{proof}, \emph{argument} (a sketch whose general validity is not established), \emph{grid} (exhaustive on the pre-registered grid), \emph{numerics}---and its provenance: pre-registered prediction, \emph{cross-check} (sealed, not pre-registered), \emph{reading} (sealed, without registered expectation) or \emph{post hoc} (after sealing, unsealed). Results of Ref.~\cite{Wang2026paper3} that are the case $N=2$ of ours are re-derived, not re-claimed.

\begin{table*}[!t]
\caption{Results and their evidence (labels as in Sec.~\ref{sec:intro}; grids in Appendix~\ref{app:grids}). ``pred.'': pre-registered prediction (stage, hypothesis); ``cross'': sealed cross-check. Check counts and details: Appendix~\ref{app:evidence}.}
\label{tab:scope}
\centering\scriptsize\setlength{\tabcolsep}{3pt}
\begin{tabular}{lp{7.0cm}ll}
\toprule
result & content & support & provenance\\
\midrule
Thm.~\ref{thm:class} & commuting gluings $=$ $u$-family; odd $L_x$ iff $u^2\equiv-1$ & proof; enumeration; grid & pred.\ E-1 H1, H2\\
Thm.~\ref{thm:deg}(i,\,iii) & $|\cR|$; frustration iff $L_x$ odd and $N\equiv2\pmod4$; Lemma~\ref{lem:sign} & proof; grid & pred.\ E-1 H3, H4; Lemma~\ref{lem:sign} post hoc\\
Thm.~\ref{thm:deg}(ii,\,iii) & seam action $P$; $\Pe^2=C$; $|\cR|=|\ker(1-P)|$; phases & proof; sealed property checks & cross (not pre-registered)\\
Thm.~\ref{thm:anatomy}, \ref{thm:lock} & literal gluing: clock rings, locking rule, spectrum, $E_0$, $\GSD$ & proof; grid; numerics & pred.\ E-2 H1--H4 (reduced coverage)\\
Prop.~\ref{prop:sigma} & row-exchanging reflection $\sigma$; Kramers--Wannier action & proof; grid & pred.\ E-2 H5\\
Prop.~\ref{prop:L} & logical module $=$ twisted homology (edge / site register) & proof / grid, argument & pred.\ E-3 H1\\
Sec.~\ref{sec:codesK} & logical operators; distances (at the listed sizes) & grid & pred.\ E-3 H2, H3\\
Prop.~\ref{prop:gates} & admissible dressed isometries, logical actions, gate groups & argument; grid & pred.\ E-3 H4, H5\\
Thm.~\ref{thm:homology} & $\Sigma_{g,k}$: consistency, $|H_1|$, block quotient, direct sums & proof; grid & pred.\ E-4 H1; other structures: reading\\
Remark~\ref{rem:fusion} & $C$-defect fusion count consistent with $|H_1|$ ($N$ odd) & arithmetic consistency & pred.\ E-4 H2\\
Prop.~\ref{prop:mono} & double-winding closure counts; monodromy multiplier $h$ & grid & pred.\ E-4 H3; $h$: reading\\
Cor.~\ref{cor:nogo} & no non-stabilizer resource from cross-caps within $\Op$ & proof (Gottesman--Knill) & not computed\\
Prop.~\ref{prop:wigner} & Wigner minima; outcome probabilities of three measurements & proof; grid & pred.\ E-4 H4\\
Sec.~\ref{sec:sym} & no protection from non-orientability itself & hand assessment & not computed\\
\bottomrule
\end{tabular}
\end{table*}

\section{Model and conventions}\label{sec:setup}

\paragraph{Weyl operators.} With $\omega=e^{2\pi i/N}$, $X|j\rangle=|j+1\rangle$ and $Z|j\rangle=\omega^j|j\rangle$, so $ZX=\omega XZ$. Weyl strings $W(v)=\prod_sX_s^{v^X_s}Z_s^{v^Z_s}$, $v\in\Z_N^{2n}$ (up to phases, which all computations track exactly), satisfy $W(v)W(w)=\omega^{\langle v,w\rangle}W(w)W(v)$ with $\langle v,w\rangle=\sum_s(v^Z_sw^X_s-v^X_sw^Z_s)$.

\paragraph{Site register.} A qudit sits on every site $(x,y)$, $x\in\Z_{L_x}$, $y=0,\dots,L_y-1$. The plaquette $p=(x,y)$, with corners SW $=(x,y)$, SE $=(x+1,y)$, NE $=(x+1,y+1)$, NW $=(x,y+1)$, carries
\begin{equation}
F_p=X_{\mathrm{SW}}Z_{\mathrm{SE}}X^\dagger_{\mathrm{NE}}Z^\dagger_{\mathrm{NW}},\qquad H=-\tfrac12\sum_p\big(F_p+F_p^\dagger\big),
\label{eq:model}
\end{equation}
the $\Z_N$ plaquette model \cite{YouWen2012}; for $N=2$ it is Wen's model, $H=-\sum_pF_p$ \cite{Wen2003plaquette,Wang2026paper3}. Plaquettes $(x,y)$ and sites $(x,y)$ have colour A ($x+y$ even) or B (odd). A site receives its $X$-type factors from the two plaquettes of its own colour and its $Z$-type factors from the other two. Conjugating the odd sites by the Fourier transform $\cF$ ($\cF X\cF^\dagger=Z$, $\cF Z\cF^\dagger=X^\dagger$) makes the A plaquettes $X$-type and the B plaquettes $Z$-type: on a torus with even sides Eq.~\eqref{eq:model} is the $\Z_N$ toric code of the lattice whose vertices, edges and faces are the A plaquettes, the sites and the B plaquettes. A plaquette with eigenvalue $\omega^k$ carries charge $k$; the charges of the two colours are the anyons $e$ and $m$, with labels $(A,B)\in\Z_N^2$, braiding phase $\omega^{AB'+A'B}$ and spin $\omega^{AB}$; a one-site translation exchanges them \cite{Wen2003plaquette}.

\paragraph{Edge register (control).} Qudits sit on the oriented edges $h(x,y)\colon(x,y)\to(x+1,y)$ and $v(x,y)\colon(x,y)\to(x,y+1)$; $A_v$ carries $X$ on the edges leaving $v$ and $X^\dagger$ on those entering it, $B_p$ carries $Z^{\pm1}$ along the counterclockwise boundary ($+1$ along the arrow). This is the $\Z_N$ toric code \cite{Kitaev2003}, the homological code of the cellulation.

\begin{figure*}[!t]
\centering\includegraphics[width=0.98\textwidth]{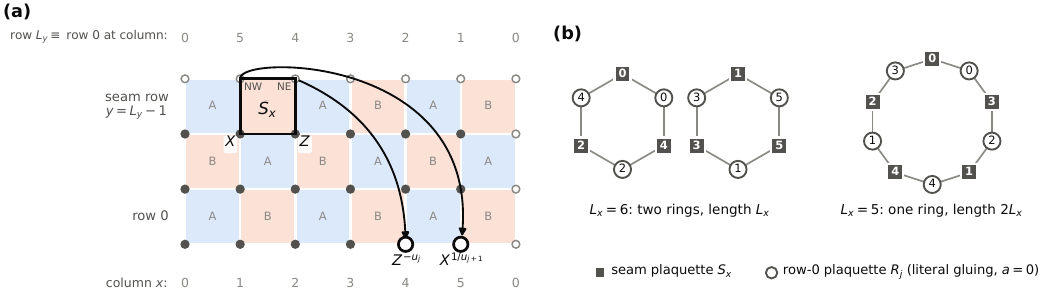}
\caption{(a) Mirror Klein bottle of the site register ($L_x=6$, $L_y=3$, $a=0$; colours A/B). The virtual row $L_y$ is row 0 read at column $a-x$ (top labels). A seam plaquette $S_x$ keeps $X$, $Z$ on its near corners; its far corners land on row 0 at columns $j=a-x-1$ and $j+1$, where the $u$-family places $Z^{-u_j}$ and $X^{1/u_{j+1}}$ (Theorem~\ref{thm:class}). (b) Non-commutation graph of the literal gluing ($a=0$): $S_x$ (squares) and $R_j$ (circles), labelled by column; $S_x$ meets $R_{a-x-2}$ with phase $\omega$ and $R_{a-x}$ with $\omega^{-1}$, giving two rings of length $L_x$ (even $L_x$) or one of length $2L_x$ (odd $L_x$) (Theorem~\ref{thm:anatomy}). For $N=2$ these are the graphs of Ref.~\cite{Wang2026paper3}, Fig.~2(a,b).}
\label{fig:seam}
\end{figure*}

\paragraph{Mirror Klein bottle.} Row $L_y$ is identified with row $0$ by $x\mapsto a-x$ (offset $a$). In the edge register the vertical edge leaving $(x,L_y-1)$ enters the vertex $(a-x,0)$, and the top edge of the plaquette $(x,L_y-1)$, traversed counterclockwise, is the row-0 edge $h(a-x-1,0)$ traversed along its arrow, so the seam-row $B_p$ carries $Z$ there instead of $Z^\dagger$. The code is again a homological code of a cellulation of the Klein bottle $K$, and the seam reverses the sign of fluxes: $P=\mathrm{diag}(1,-1)$ on $(e,m)$. In the site register the seam plaquettes $S_x:=F_{(x,L_y-1)}$ have their far corners on row 0, NE at column $j=a-x-1$ and NW at column $j+1$ (Fig.~\ref{fig:seam}a); $R_j:=F_{(j,0)}$ are the row-0 plaquettes, all others are bulk plaquettes, and there are $N_p=L_xL_y=n$ plaquettes and qudits. The reflection reverses the pattern $X,Z,X^\dagger,Z^\dagger$, so the far corners must be chosen:

\begin{definition}[local monomial far-corner gluing]\label{def:gluing}
Assign $(p_j,q_j,p'_j,q'_j)\in\Z_N^4$ to every column $j$ of row 0. In $S_x$ the NE far corner (column $j=a-x-1$) carries $X^{-p_j}Z^{-q_j}$ and the NW far corner (column $j+1$) carries $X^{-p'_{j+1}}Z^{-q'_{j+1}}$; everything else is as in Eq.~\eqref{eq:model}. The \emph{literal} gluing has $(p_j,q_j)=(1,0)$, $(p'_j,q'_j)=(0,1)$: the far corners keep $X^\dagger$ and $Z^\dagger$.
\end{definition}
Phases on the monomials affect only the scalars of relations [Theorem~\ref{thm:deg}(iii)]. By Theorem~\ref{thm:class} the commuting gluings are the \emph{$u$-gluings}: the NE and NW far corners on column $j$ carry $Z^{-u_j}$ and $X^{1/u_j}$, with $u_j=u$ for even and $-1/u$ for odd $j$; $u=\pm1$ is the Fourier dressing ($\cF$, $\cF^\dagger$ alternating by column), and $N=2$, $u=1$ is the Hadamard-dressed gluing of Ref.~\cite{Wang2026paper3}. For even $L_x$ there are two branches. In the \emph{diagonal branch} ($a+L_y$ odd) every site still receives its $X$-type factors from plaquettes of its own colour: the colouring continues across the seam, which maps each anyon type to itself. In the \emph{exchange branch} ($a+L_y$ even) the colours, and with them $e$ and $m$, are exchanged across the seam. For odd $L_x$ the colouring is inconsistent around $x$ and there is one charge type. ``Grid'' statements hold on the pre-registered grids (Appendix~\ref{app:grids}) and are not extrapolated.

\section{Local gluings: the \texorpdfstring{$u$}{u}-family}\label{sec:class}

\begin{theorem}[local gluings]\label{thm:class}
Let $N\ge2$, $L_x\ge3$, $L_y\ge3$. A local monomial far-corner gluing makes all plaquettes commute if and only if, for every column $j$,
\begin{equation}
p_j=q'_j=0,\qquad q_jp'_j\equiv-1,\qquad p'_{j+1}\equiv q_j\pmod N .
\label{eq:conditions}
\end{equation}
Equivalently, the NE and NW far corners on column $j$ carry $Z^{-u_j}$ and $X^{1/u_j}$ with $u_j\in\Z_N^\times$ and $u_{j+1}=-1/u_j$, so $u_j=u$ for even and $-1/u$ for odd $j$. For even $L_x$ every unit $u$ gives a commuting gluing ($|\Z_N^\times|$ of them); for odd $L_x$ the chain closes iff $u^2\equiv-1$. Hence commuting gluings with odd $L_x$ exist iff $-1$ is a square mod $N$, i.e.\ iff $4\nmid N$ and every odd prime factor of $N$ is $\equiv1\pmod4$.
\end{theorem}
\begin{proof}
$S_x$ shares far-corner sites only with $R_{j-1},R_j,R_{j+1}$ ($j=a-x-1$; distinct as $L_x\ge3$) and with $S_{x\pm1}$, whose far corners sit on columns $j\pm1$; all other overlaps are those of the torus and cancel ($L_y\ge3$ keeps the row-0 plaquettes away from the seam row). Writing $AB=\omega^{e(A,B)}BA$, $e(S_x,R_{j-1})=p_j$, $e(S_x,R_{j+1})=-q'_{j+1}$, $e(S_x,R_j)=p'_{j+1}-q_j$, and $e(S_{x-1},S_x)=1+q_{j+1}p'_{j+1}-p_{j+1}q'_{j+1}$ (one near and one far corner in common). Setting them to zero gives Eq.~\eqref{eq:conditions}; then $u_j:=q_j$ is a unit and $u_{j+1}=-1/u_j$. The recursion has period two, so it closes on even cycles for every $u$ and on odd ones iff $u=-1/u$. The last statement is the classical criterion for $x^2\equiv-1\pmod N$.
\end{proof}

\smallskip
For even $L_x$ every unit works, and for odd $L_x$ only the closing roots $u^2\equiv-1$; locally every unit is allowed because the $u$-gluing is the quotient by the glide $(x,y)\mapsto(a-x,y+L_y)$ dressed with $X\mapsto Z^{-u_s}$, $Z\mapsto X^{1/u_s}$ ($u_s=u$ or $-1/u$ by sublattice), which maps every plaquette to some $F_{p'}^{\pm u^{\pm1}}$ (derivation; checked post hoc): a symmetry of the stabilizer group for every $u$, of $H$ only for $u\equiv\pm1$. The arithmetic condition is the integer set of the time-reversal criterion for $U(1)_k$ Chern--Simons theory \cite{DelmastroGomis2021}; specific here is its role as a criterion for local lattice gluings.

\begin{theorem}[seam action and code dimension]\label{thm:deg}
For a $u$-gluing with $L_x,L_y\ge3$, let $\cR=\{(\kappa_p)\in\Z_N^{N_p}:\prod_pF_p^{\kappa_p}\propto1\}$ be the relation module and $\gamma_\pm=\gcd(u\pm1,N)$.
\begin{enumerate}
\item[(i)] $|\cR|=\gamma_-\gamma_+$ in the diagonal branch, and $|\cR|=\gcd(2,N)$ in the exchange branch and for odd $L_x$.
\item[(ii)] A charge carried upward through the seam has its label $(A,B)$ mapped, in the diagonal and the exchange branch respectively, by
\begin{equation}
\Pd=\begin{pmatrix}-u&0\\0&u^{-1}\end{pmatrix},\qquad \Pe=\begin{pmatrix}0&-u\\u^{-1}&0\end{pmatrix},
\label{eq:P}
\end{equation}
and for odd $L_x$ (one charge type) by $-u$ with $u^2=-1$. Both matrices invert the braiding phase and the spin, $\Pe^2=-1=C$, and $|\cR|=|\ker(1-P)|$, the number of $P$-invariant labels, in every case.
\item[(iii)] With phase-free monomials all relation scalars are $1$, except for odd $L_x$ and $N\equiv2\pmod4$, where $(\prod_pF_p)^{N/2}=-1$ and no state satisfies all plaquettes. Phases change only the scalars: multiplying one seam plaquette by $\omega$ leaves $\cR$ unchanged and, for even $N$ in the exchange branch or with odd $L_x$, where $\cR$ is generated by $(\prod_pF_p)^{N/2}$, flips its scalar. The code dimension is $|\cR|$ whenever all scalars are $1$; only $|\cR|$ is an invariant of the gluing.
\end{enumerate}
\end{theorem}

\begin{lemma}[sign of the dressed product]\label{lem:sign}
For $N\equiv2\pmod4$ and every $u$-gluing, $\prod_pF_p^{N/2}=(-1)^{L_x}$.
\end{lemma}
\begin{proof}
$\tilde X=X^{N/2}$ and $\tilde Z=Z^{N/2}$ satisfy $\tilde X^2=\tilde Z^2=1$ and $\tilde Z\tilde X=\omega^{N^2/4}\tilde X\tilde Z=-\tilde X\tilde Z$ ($N/2$ odd), the qubit Pauli relations. $F_p^{N/2}=\tilde X_{\rm SW}\tilde Z_{\rm SE}\tilde X_{\rm NE}\tilde Z_{\rm NW}$, and as every $u_j$ is odd the far corners become $\tilde Z$ (NE) and $\tilde X$ (NW): the product is that of the $N=2$ Hadamard-dressed gluing. Order its plaquettes by $p=yL_x+x$ and multiply site by site. A site with $y\ge1$ receives its four factors in the order $XZZX$ ($x\ge1$) or $ZXXZ$ ($x=0$): product $+1$. Row-0 site $j$ receives $(Z,X)$ from $R_{j-1},R_j$, reversed at $j=0$, followed by $(Z,X)$ from the far corners of $S_{a-j-1},S_{a-j}$, reversed at $j\equiv a$; two pairs in the same order give $XZXZ$ or $ZXZX=-1$, in opposite orders $+1$. So the sign is $-1$ at the $L_x-2$ sites $j\notin\{0,a\}$ if $a\not\equiv0$, and at all $L_x$ sites if $a\equiv0$; either way $(-1)^{L_x}$.
\end{proof}
For $N=2$ this sign was only observed, at the sizes computed, in Ref.~\cite{Wang2026paper3}; the proof above was supplied by the internal review of draft v0.1 and checked post hoc.

\begin{proof}[Proof of Theorem~\ref{thm:deg}]
(i) $\prod_pF_p^{\kappa_p}\propto1$ iff the exponents vanish on every site. A site with $y\ge1$ is SW and NE corner of $(x,y)$ and $(x-1,y-1)$ and SE and NW corner of $(x-1,y)$ and $(x,y-1)$, so $\kappa$ is constant on colour classes, $\kappa_{\rm A}$ and $\kappa_{\rm B}$ ($\kappa_{\rm A}=\kappa_{\rm B}$ for odd $L_x$). Row-0 site $j$ gives $\kappa(R_j)+u_j^{-1}\kappa(S_{a-j})=0$ and $\kappa(R_{j-1})-u_j\kappa(S_{a-j-1})=0$, and $S_{a-j}$, $R_j$ have the same colour iff $a+L_y$ is odd. Hence $(u+1)\kappa_{\rm A}=(u-1)\kappa_{\rm B}=0$ (diagonal), $\kappa_{\rm B}=u\kappa_{\rm A}$ and $2\kappa_{\rm A}=0$ (exchange), $(u\pm1)\kappa=0$ (odd $L_x$), with no division by non-units. (ii) (Derivation.) A $Z^q$ at the last site $(x,L_y-1)$ of a string along $(1,1)$ is continued through the NW far corner of $S_x$ by $Z^{-u_{a-x}q}$ on row 0; an $X^q$ string is continued through the NE far corner by $X^{q/u_{a-x-1}}$. Each string carries one colour, so a label continued on colour A is multiplied by $-u$ and one continued on colour B by $u^{-1}$, colours being preserved (diagonal) or exchanged (exchange branch): Eq.~\eqref{eq:P}, up to a sign convention for labels. The entries of $P$ multiply to $-1$, which inverts $AB'+A'B$ and $AB$, and the equations of (i) are the fixed-point equations of $P$. (iii) A relation supported on one colour has only $X$-type or only $Z$-type factors on each site and scalar $1$; in the diagonal branch these generate $\cR$. Otherwise, for even $N$, $\cR$ is generated by $(\prod_pF_p)^{N/2}$, whose scalar is $1$ for $N\equiv0\pmod4$ ($X^{N/2}$ and $Z^{N/2}$ commute) and $(-1)^{L_x}$ for $N\equiv2\pmod4$ (Lemma~\ref{lem:sign}). Multiplying $S_x$ by $\omega$ multiplies that scalar by $\omega^{N/2}=-1$.
\end{proof}

The diagonal branch with $u=\pm1$ is an order-two parity action, with $|\cR|=N\gcd(2,N)$ as for the edge register; the exchange branch realizes an order-four seam, $\Pe^2=C$ ($N>2$), and encodes nothing for odd $N$. For odd $L_x$ the $N=2$ case is the sign-frustrated dressed gluing of Ref.~\cite{Wang2026paper3}; on the grid the frustration recurs for $N=10,26$, and the codes are one-dimensional for $N=5,13,17,25$.

\section{The obstructed seam: locked clock rings}\label{sec:seam}

\begin{theorem}[obstruction of the literal gluing]\label{thm:anatomy}
Let $N\ge2$, $L_x\ge3$, $L_y\ge2$, any $a$. In the literal gluing the non-commuting pairs are exactly $\{S_x,R_{a-x-2}\}$ and $\{S_x,R_{a-x}\}$, with $S_xR_{a-x-2}=\omega R_{a-x-2}S_x$ and $S_xR_{a-x}=\omega^{-1}R_{a-x}S_x$. They form two cycles of length $L_x$ for even $L_x$ and one of length $2L_x$ for odd $L_x$ (Fig.~\ref{fig:seam}b). Along $S_x\to R_{a-x-2}\to S_{x+2}\to\cdots$ consecutive operators satisfy $U_iU_{i+1}=\omega U_{i+1}U_i$ with no operator inverted, so each cycle is a self-dual $\Z_N$ clock ring of $m=L_x/2$ (even $L_x$) or $m=L_x$ (odd $L_x$) sites, with the seam plaquettes as field terms and the row-0 plaquettes as bond terms.
\end{theorem}
\begin{proof}
With $j=a-x-1$, $S_x$ carries $X^\dagger$ on column $j$ and $Z^\dagger$ on $j+1$ of row 0, and $R_{j'}$ carries $X$ on $j'$, $Z$ on $j'+1$. For $j'=j-1$ the common site contributes $+1$, for $j'=j+1$ it contributes $-1$, and for $j'=j$ both common sites carry commuting factors; other overlaps cancel as on the torus (also for $L_y=2$). A step along the cycle advances the column by two, preserving its parity for even $L_x$ and visiting all columns for odd $L_x$. $U_{2i}=X_i$, $U_{2i+1}=Z_i^\dagger Z_{i+1}$ represents the ring.
\end{proof}

Let $D_r$ and $E_r$ be the products of the seam and of the row-0 plaquettes of ring $r$ (ring charge and ring twist); let $H_{\mathrm{clock}}(m;t,q)=-\tfrac12\sum_i(X_i+X_i^\dagger)-\tfrac12\sum_i(\omega^{t[i=m-1]}Z_i^\dagger Z_{i+1}+\mathrm{h.c.})$ on $m$ sites mod $m$, restricted to $\prod_iX_i=\omega^q$, with lowest eigenvalue $e_0(m)$ at $(t,q)=(0,0)$; and let $\delta=2$ if $L_x$ and $a+L_y$ are both even, $\delta=1$ otherwise.

\begin{theorem}[locking rule and exact reduction]\label{thm:lock}
For the literal gluing with $N\ge2$, $L_x\ge3$, $L_y\ge2$:
(a) the relations form $\Z_N^\delta$, generated by the product of all plaquettes and, for $\delta=2$, the product of the colour-A plaquettes, which commute pairwise; both scalars are $1$, the product of all plaquettes taken in the order $S_0,\dots,S_{L_x-1},R_0,\dots,R_{L_x-1}$, bulk (the index order gives the same);
(b) the centre is generated by the bulk plaquettes $F_b$, the ring charges $D_r$ and the ring twists $E_r$; with eigenvalues $\omega^{f_b}$, $\omega^{q_r}$, $\omega^{t_r}$, the realized joint eigenvalues are exactly those with
\begin{equation}
\textstyle\sum_bf_b+\sum_r(t_r+q_r)\equiv0\pmod N
\label{eq:lock}
\end{equation}
and, for $\delta=2$, $\sum_{b\in\mathrm A}f_b+q_r+t_{r'}\equiv0$, where $r$ is the ring whose seam plaquettes have colour A and $r'$ the other; each realized joint eigenspace has dimension $N^{2L_x-4+\delta}$ (even $L_x$) or $N^{2L_x-1}$ (odd $L_x$), in which each ring carries one irreducible representation of dimension $N^{m-1}$ with multiplicity $\nu=N^{L_x-2+\delta}$ (even $L_x$) or $N^{L_x}$ (odd $L_x$), independent of $L_y$ except through the parity of $a+L_y$;
(c) every eigenvalue of $H$ is $-\sum_b\cos(2\pi f_b/N)+\sum_re_r$, $e_r\in\mathrm{spec}\,H_{\mathrm{clock}}(m_r;t_r,q_r)$, over the allowed labels, with multiplicity $\nu$ times the ring multiplicities;
(d) the ground states have all $f_b=0$ and every ring in $(t,q)=(0,0)$:
\begin{equation}
E_0=-(N_p-2L_x)+\textstyle\sum_re_0(m_r),\qquad \GSD=\nu .
\label{eq:E0}
\end{equation}
For $N=2$ this is the result of Ref.~\cite{Wang2026paper3}; for $N=4$ a ring is two Ising chains and $e_0(m)=-2/\sin(\pi/2m)$ as for $N=2$, so every lattice has the same $E_0$ for $N=4$ as for $N=2$. For $N=3$ and $N\ge5$ we give no closed form for $e_0(m)$ and use the numerical ground energy of the ring.
\end{theorem}
\begin{proof}
(a) Each site carries one $X$, $X^\dagger$, $Z$, $Z^\dagger$, so relation coefficients are constant on colour classes; the seam identifies $\kappa(R_j)$ with $\kappa(S_{a-j-1})$, which crosses colours unless $a+L_y$ is even; nothing is divided, so composite $N$ adds no torsion. In the stated order every site contributes the scalar $1$ except, for $a\not\equiv0$, two row-0 sites contributing $\omega$ and $\omega^{-1}$: the cancellation is global, not site by site. For $\delta=2$ the non-commuting pairs join opposite colours, and each site carries $X$ and $X^\dagger$, or $Z$ and $Z^\dagger$, from the colour-A plaquettes. (b), (c): the cycle algebra has a unique irreducible representation, of dimension $N^{m-1}$, for each central character; count dimensions. (d) For fixed $t$, $H_{\mathrm{clock}}(m;t)$ on all $N^m$ states has non-positive off-diagonal elements in the $Z$ basis and is irreducible, so by Perron--Frobenius its ground state is unique, positive, and has $q=0$. The ring translation exchanges $D$ and $E$, so $\mathrm{spec}(t,q)=\mathrm{spec}(q,t)$ and $\min\mathrm{spec}(t,0)>\min\mathrm{spec}(0,0)$ for $t\neq0$. For $N=4$, a Gray-code map writes the ring as two critical Ising chains.
\end{proof}

\begin{table}[!t]
\caption{Ground energies of the literal gluing, Eq.~\eqref{eq:E0}, against exact diagonalization on 9 of the 17 lattices of E-2 (dev.\ = largest deviation over $a=0,1$). $\GSD$ ($a=0/1$): counted from complete spectra (d: dense; b: block-resolved by bulk labels), or $\nu$ of Theorem~\ref{thm:lock} in italics where Lanczos does not resolve degeneracies (L: with low levels; g: ground energy only).}
\label{tab:E0}
\centering\scriptsize\setlength{\tabcolsep}{2.5pt}
\begin{tabular}{rlrrrrl}
\toprule
$N$ & size & states & $E_0$ & dev. & $\GSD$ & \\
\midrule
2 & $4\times3$ & 4\,096 & $-9.65685425$ & $7\times10^{-14}$ & $8/16$ & d\\
3 & $4\times2$ & 6\,561 & $-5.46410162$ & $7\times10^{-14}$ & $81/27$ & d\\
4 & $3\times2$ & 4\,096 & $-4.00000000$ & $5\times10^{-14}$ & $64/64$ & d\\
3 & $3\times3$ & 19\,683 & $-6.84232922$ & $5\times10^{-15}$ & $27/27$ & b, L\\
3 & $4\times3$ & 531\,441 & $-9.46410162$ & $3\times10^{-14}$ & \textit{27/81} & L\\
5 & $3\times3$ & 1\,953\,125 & $-7.23525950$ & $1\times10^{-14}$ & \textit{125/125} & L\\
3 & $5\times3$ & 14\,348\,907 & $-11.20032229$ & $5\times10^{-14}$ & \textit{243/243} & g\\
4 & $4\times3$ & 16\,777\,216 & $-9.65685425$ & $6\times10^{-14}$ & \textit{64/256} & g\\
6 & $3\times3$ & 10\,077\,696 & $-7.45670722$ & $1\times10^{-13}$ & \textit{216/216} & g\\
\bottomrule
\end{tabular}
\end{table}

\begin{proposition}[row-exchanging reflection]\label{prop:sigma}
The site permutation $\sigma$: $(x,0)\mapsto(x,0)$, $(x,y)\mapsto(a-x,L_y-y)$ for $y\ge1$, implemented by the unitary $U_\sigma$, satisfies $U_\sigma F_pU_\sigma^\dagger=F_{\sigma(p)}^\dagger$, is a symmetry of $H$, and exchanges the seam row and row 0 ($R_j\leftrightarrow S_{a-j-1}$ up to inverses), hence the field and bond terms of the rings: for odd $L_x$ it acts on the ring as $U_i\mapsto U_{i+L_x}^\dagger$, a Kramers--Wannier duality followed by a translation; for even $L_x$ it exchanges the two rings. A uniform detuning of the seam row against row 0 breaks $\sigma$.
\end{proposition}
\begin{proof}
Lift to the double cover, where $\sigma$ is $y\mapsto-y$; bulk plaquettes map as $(x,y)\mapsto(a-x-1,L_y-1-y)$.
\end{proof}
For reweighted plaquette terms, invariance under $\sigma$ and the half translation forces both ring masses to vanish for $L_x\equiv2\pmod4$; for $L_x\equiv0\pmod4$ a staggered reweighting invariant under $\sigma$, translation and $C$ gives the rings opposite non-zero masses.

Self-dual clock rings are known: three-state Potts criticality for $N=3$, two Ising chains for $N=4$, and a critical phase containing the self-dual point for $N\ge5$ \cite{LiYangTuCheng2015,SunVekuaCobaneraOrtiz2019}. As a pre-registered anchor, the $N=3$ ring gives the ratio of the lowest charged-sector gap ($q=1$) to the first neutral gap, $\Delta_\sigma/\Delta_\varepsilon=0.16200$ at $m=12$ ($0.16298$ extrapolated in $m^{-2}$; Potts ratio of scaling dimensions $x_\sigma/x_\varepsilon=1/6$), and central charge $c=0.7713$ from the registered fit, with the velocity $1.3478$ inferred from $\Delta_\varepsilon$ assuming $x_\varepsilon=4/5$; both are within the registered $\pm10\%$, and the exact velocity $3\sqrt3/4$ gives $c=0.8002$ (reading). As for $N=2$, the ring degeneracy belongs to the frustrated Hamiltonian; we make no protection claim.

\section{Arithmetic and codes}\label{sec:codes}

\subsection{Klein-bottle codes}\label{sec:codesK}

For a stabilizer group $\cS$, $\cS^\perp$ denotes the Weyl operators commuting with $\cS$ (modulo phases) and $\cL=\cS^\perp/\cS$ the logical module; the code dimension is $|\cL|^{1/2}$.

\begin{proposition}[logical module]\label{prop:L}
For the $u$-gluing with even $L_x$, on the registered grid,
\begin{equation}
\cL\cong H_1(K;\Z_N^2{}_\rho)\cong\coker(1+P)\oplus\ker(1-P),
\label{eq:LK}
\end{equation}
with $\rho$ trivial along rows and $\rho=P$ across the seam: $(\Z_{\gamma_-}\oplus\Z_{\gamma_+})^2$ in the diagonal branch and $(\Z_{\gcd(2,N)})^2$ in the exchange branch. The edge register is the diagonal case $u=-1$, $P=\mathrm{diag}(1,-1)$, and $\cL\cong(\Z_N\oplus\Z_{\gcd(2,N)})^2$ for every size.
\end{proposition}
\noindent\emph{Proof for the edge register; argument for the site register.} In the edge register $\cS^\perp$ consists of $X$-type cocycles and $Z$-type cycles and $\cS$ of coboundaries and boundaries, so $\cL\cong H^1(K;\Z_N)\oplus H_1(K;\Z_N)\cong(\Z_N\oplus\Z_{\gcd(2,N)})^2$. In the site register, string operators modulo stabilizers are 1-chains whose label coefficients are twisted by $P$ across the seam; for $\pi_1(K)=\langle\alpha,\beta\,|\,\beta\alpha\beta^{-1}\alpha\rangle$, $\rho(\alpha)=1$, $\rho(\beta)=P$, Fox calculus gives $\partial_2=(1+P,0)$, $\partial_1=(0,P-1)$, hence Eq.~\eqref{eq:LK}. In the diagonal branch, conjugating the odd sites by $\cF$ gives a CSS code with the A and B plaquettes as checks and unit coefficients, i.e.\ $H^1\oplus H_1$ of a cellulation of $K$ with unit-twisted incidence numbers; that this complex (or its exchange-branch analogue) is the twisted complex of Eq.~\eqref{eq:LK} is not proved here.

\smallskip
For composite $N$, $\cL$ is in general not the square of a cyclic group (Table~\ref{tab:L}): at $N=8$, $u=\pm1$ give $(\Z_2\oplus\Z_8)^2$ and $u=\pm3$ give $(\Z_2\oplus\Z_4)^2$; at $N=12$ all four units give $(\Z_2\oplus\Z_{12})^2$, split differently between the colours. (At $N=15$, $u=4$ gives $(\gamma_-,\gamma_+)=(3,5)$ and $(\Z_3\oplus\Z_5)^2\cong\Z_{15}^2$.)

\paragraph{Logical operators} (grid). The row loop $\Lambda_y(p,q)$ places $X^pZ^q$ on the even and $X^qZ^p$ on the odd sites of row $y$; its label is $q$ on colour $y$ and $p$ on colour $y+1$. In the diagonal branch it commutes with $\cS$, lies in $\cS$ iff its label $(A,B)$ has $\gamma_-|A$, $\gamma_+|B$, and $\Lambda_y(p,q)\equiv\Lambda_{y+1}(q,p)$ mod $\cS$; $h_{\rm A}=[\Lambda_1(1,0)]$ and $h_{\rm B}=[\Lambda_1(0,1)]$ have orders $\gamma_-$, $\gamma_+$. The axis diagonals $\prod_yZ^q_{(x_0+y,y)}$, $2x_0\equiv a-L_y+1$, and $\prod_yX^p_{(x_1-y,y)}$, $2x_1\equiv a+L_y-1$, close iff $(1+u_{x_0})q\equiv0$, resp.\ $(u_{x_1}-1)p\equiv0$, and are then non-trivial modulo $\cS$ and the row loops. Each colour has a one-site-per-row vertical representative, with two turns where the straight diagonals all carry the other colour ($L_x/2$, $L_y$ even; 74 codes). In the exchange branch with $N$ even, $\Lambda_y(p,q)\in\cS$ iff $p+q$ is even, and for even $a$ the columns $\prod_y(X^{N/2}Z^{N/2})_{(c,y)}$, $c\in\{a/2,a/2+L_x/2\}$, are non-trivial and anticommute with $\Lambda_y(1,0)$. With the four edge-register loop types, these and $\cS$ generate $\cS^\perp$ on every grid code.

\paragraph{Distances} (grid only; site register with $u=\pm1$, $N\le8$, and $N=8$, $u=3$ in the diagonal branch, $u=1$ and even $N$ in the exchange branch). Diagonal branch (dimension $>1$): $d=\min(L_x,L_y)$; exchange branch: $\min(L_x,2\lceil L_y/2\rceil)$; edge register, $N$ even: $d_X=\min(L_x,L_y+[a\text{ even}])$, $d_Z=\min(L_x,L_y+[a\text{ odd}])$ as for $N=2$; $N$ odd: $d_Z=L_y+[a\text{ odd}]$ and $d_X=L_x$ at the computed sizes. The last fails off the grid: an $X$-type logical made of two dual cross-cap cores and a seam-row string has weight 13 at $N=3,5$, $14\times3$, and 14 at $N=3$, $16\times3$ ($a=1$; readings).

\begin{table}[!t]
\caption{Site codes at $4\times3$, $a=0$ (diagonal branch; E-3): logical module $\cL$, dimension $|\cL|^{1/2}$, order $|\Gamma|$ of the group generated by the logical actions of all dressed isometries modulo logical Weyl operators ($N=15$ is not on the isometry grid), and whether the Dehn twist acts ($\iota\neq0$). Edge register, $4\times3$, $a=0$: $|\Gamma|=2,2,4,4,4,6,8,6,8,10,8$ for $N=2,\dots,12$.}
\label{tab:L}
\centering\scriptsize\setlength{\tabcolsep}{3pt}
\begin{tabular}{rrrrlrrc}
\toprule
$N$ & $u$ & $\gamma_-$ & $\gamma_+$ & $\cL$ & dim & $|\Gamma|$ & $\iota\neq0$\\
\midrule
2 & 1 & 2 & 2 & $\Z_2^4$ & 4 & 4 & yes\\
3 & 1 & 3 & 1 & $\Z_3^2$ & 3 & 2 & no\\
4 & 1 & 4 & 2 & $(\Z_2\oplus\Z_4)^2$ & 8 & 4 & yes\\
5 & 2 & 1 & 1 & $0$ & 1 & 1 & no\\
6 & 1 & 6 & 2 & $(\Z_2\oplus\Z_6)^2$ & 12 & 4 & yes\\
8 & 1 & 8 & 2 & $(\Z_2\oplus\Z_8)^2$ & 16 & 8 & yes\\
8 & 3 & 2 & 4 & $(\Z_2\oplus\Z_4)^2$ & 8 & 2 & no\\
10 & 3 & 2 & 2 & $\Z_2^4$ & 4 & 4 & yes\\
12 & 5 & 4 & 6 & $(\Z_2\oplus\Z_{12})^2$ & 24 & 8 & yes\\
15 & 4 & 3 & 5 & $\Z_{15}^2$ & 15 & --- & ---\\
16 & 1 & 16 & 2 & $(\Z_2\oplus\Z_{16})^2$ & 32 & 16 & yes\\
16 & 7 & 2 & 8 & $(\Z_2\oplus\Z_8)^2$ & 16 & 4 & no\\
\bottomrule
\end{tabular}
\end{table}

\subsection{Lattice isometries and gates}\label{sec:gates}

The isometries of the lattice Klein bottle are the maps $(x,y)\mapsto(s_xx+c,\,s_yy+c_y)$, $s_x,s_y=\pm1$, that normalize the deck group generated by $(x,y)\mapsto(x+L_x,y)$ and $(x,y)\mapsto(a-x,y+L_y)$: $c\in\{0,L_x/2\}$ for $s_x=1$, $c\in\{a,a+L_x/2\}$ for $s_x=-1$, $c_y\in\Z_{L_y}$; $8L_y$ maps in all \cite{Wang2026paper3}. The \emph{twist} $\tau=1$ iff $c$ is the half-period choice (the half translation is not isotopic to the identity and acts on homology as a Dehn twist \cite{Wang2026paper3}); the \emph{colour parity} $s=c+c_y-[s_x=-1]-[s_y=-1]\bmod2$ is the change of plaquette colour under the map. A \emph{dressed automorphism} is the induced site permutation followed by single-site Cliffords mapping each plaquette to a unit power of the plaquette at the image position; such dressings are monomial. The multiplier $\sM_\lambda$ ($\lambda\in\Z_N^\times$) acts as $X\mapsto X^\lambda$, $Z\mapsto Z^{1/\lambda}$ on even and with $\lambda\leftrightarrow1/\lambda$ on odd sites. In the diagonal branch we use the basis $(h_{\rm A},h_{\rm B},v_{\rm A},v_{\rm B})$ of $\cL$: the row loops above and the classes of one-site-per-row vertical loops of colour A and B whose labels $\ell_{\rm A},\ell_{\rm B}$ generate the two components of $\ker(1-P)$ (orders $\gamma_+$, $\gamma_-$); the edge register uses its $e$ and $m$ loops in the same roles. Modulo logical Weyl operators $\sM_\lambda$ acts as $\mathrm{diag}(\lambda^{-1},\lambda,\lambda^{-1},\lambda)$. Let $\iota:\ker(1-P)\to\coker(1+P)$ be the natural map, sending a vertical loop to the row loop with the same label ($v_{\rm A}\mapsto(\ell_{\rm A}\bmod\gamma_-)h_{\rm A}$, $v_{\rm B}\mapsto(\ell_{\rm B}\bmod\gamma_+)h_{\rm B}$; $2\iota=0$), and $J=1+\iota$.

\begin{proposition}[gates from lattice isometries]\label{prop:gates}
On the isometry grid, within the single-site dressing class:
(a) A dressed automorphism exists iff, in the diagonal branch, ($s=0$ and ($s_y=1$ or $u^2\equiv1$)) or ($s=1$ and (($s_y=1$ and $u^2\equiv-1$) or $N=2$)); in the exchange branch, iff $c+c_y+[s_y=-1]$ is even or $-1$ is a square mod $N$; always in the edge register. The dressings of the identity are the multipliers ($\lambda\in\Z_N^\times$; $\lambda^2\equiv1$ in the exchange branch), and each admissible isometry has as many.
(b) Modulo logical Weyl operators, colour-preserving dressed automorphisms in the diagonal branch and all those of the edge register act as $\sM_\lambda J^\tau$; on codes of dimension $>1$ a non-trivial colour exchange occurs in the diagonal branch only for $u^2\equiv-1$ (on the grid: $N=2$, and $N=10$ with $u=3,7$) and acts outside $\{\sM_\lambda\}\langle J\rangle$; in the exchange branch ($N$ even) every dressed automorphism, including the geometric colour exchanges allowed by (a), is a logical Weyl operator.
(c) The generated group has order $|\Z_{\ell}^\times|\,2^{[\iota\neq0]}\,2^{[\text{colour exchange}]}$, $\ell=\mathrm{lcm}(\gamma_-,\gamma_+)$ (diagonal), and $|\Z_N^\times|\,2^{[N\text{ even}]}$ (edge), and $\iota\neq0$ iff $\gamma_-\gamma_+\nmid N$, i.e.\ iff $N$ is even and $u\equiv\pm1\pmod{2^{v}}$, $2^v$ being the largest power of 2 dividing $N$.
\end{proposition}
Necessity in (a) is an argument valid beyond the grid: the induced (anti-)automorphism $\chi$ of the labels must satisfy $\chi P\chi^{-1}=P^{s_y}$; sufficiency, (b) and (c) are exhaustive on the grid. At $N=2$ the isometries give $\mathrm{SWAP}^\tau(\Had\otimes\Had)^s$ \cite{Wang2026paper3}; for $\Z_N$ the ingredients separate. The Dehn twist acts as $v\mapsto v+\iota(v)$, a CNOT-type map on the order-two parts; for odd $N$ the half translation is a multiplier up to logical Weyl operators. A non-trivial colour exchange ($u^2\equiv-1$, dimension $>1$) forces $N\equiv2\pmod4$ and dimension $4$ for $N>2$. The multipliers exist equally on the torus, where every isometry, colour exchanges included, has a dressing (controls at $4\times4$, $6\times4$, $N\le8$): within this class the scarcity of Fourier-type gates comes from the non-orientable seam. Locality-preserving gates are constrained by anyon data and compatibility with mapping-class-group actions \cite{Beverland2016}. The Clifford restriction of Bravyi and Koenig applies to their two-dimensional topological qubit stabilizer-code setting \cite{BravyiKoenig2013}; it is not used here as a theorem for composite $N$. Proposition~\ref{prop:gates} gives the dressed-isometry criteria within its stated class and grid.

\subsection{Twisted homology of surfaces with cross-caps}\label{sec:cross}

Let $\Sigma_{g,k}$ ($k\ge1$; $\Sigma_{0,1}=\mathbb{RP}^2$) be the connected sum of $g$ tori and $k$ cross-caps, $\pi_1=\langle a_i,b_i,c_j\,|\,\mathfrak r\rangle$, $\mathfrak r=\prod_i[a_i,b_i]\prod_jc_j^2$, coefficients $\Z_N^2$, $\rho(a_i)=\rho(b_i)=1$, $\rho(c_j)=P^{\eta_j}$ with $P\in\{\Pd,\Pe\}$ and the \emph{uniform} pattern $\eta_j=1$ or the \emph{alternating} pattern $\eta_j=(-1)^{j-1}$ ($k$ even). The Klein bottle of Sec.~\ref{sec:codesK} is alternating with $k=2$ ($c_1=\alpha\beta\mapsto P$, $c_2=\beta^{-1}\mapsto P^{-1}$). Let $Q_k=\sum_{i<k}P^{2i}$.

\begin{theorem}[twisted homology]\label{thm:homology}
Let $g\ge0$, $k\ge1$ and $2g+k\ge2$.\footnote{This is the domain in the setting of the E-4 pre-registration, \texttt{prereg/e4\_prereg.md}. Its H1(d) uses $\mathbb{RP}^2$ only as an auxiliary case for handle splitting, outside the registered grid; Example~\ref{ex:rp2} treats it separately.} Parts (b)--(e) assume the consistency condition in (a).
\begin{enumerate}
\item[(a)] The local system is consistent iff $P^{2\sum_j\eta_j}=1$: for the uniform pattern iff $P^{2k}=1$ (exchange: $k$ even or $N=2$; diagonal: $u^{2k}\equiv1$); the alternating pattern always is.
\item[(b)] For the uniform pattern,
\begin{equation}
H_1\cong(\Z_N^2)^{2g}\oplus\frac{\ker\big[d_1:(\Z_N^2)^k\to\Z_N^2\big]}{\im\big[d_2:\Z_N^2\to(\Z_N^2)^k\big]},
\label{eq:block}
\end{equation}
$d_1(m_1,\dots,m_k)=\sum_j(P-1)m_j$, $d_2(m)=\big((1+P)m,P^2(1+P)m,\dots,P^{2(k-1)}(1+P)m\big)$; for the alternating pattern the $j$-th blocks carry $P^{\eta_j}$ and prefactors $P^{2(\eta_1+\dots+\eta_{j-1})}$.
\item[(c)] For both patterns
\begin{equation}
|H_1|=N^{2(2g+k-1)}\,\frac{|\ker(1+P)|}{|\im(P-1)|},
\label{eq:order}
\end{equation}
that is, $N^{2(2g+k-2)}\gcd(2,N)^2$ in the exchange branch and $N^{2(2g+k-2)}(\gamma_-\gamma_+)^2$ in the diagonal branch.
\item[(d)] $H_1\cong\Z_N^{2(2g+k-2)}\oplus\coker(1+P)\oplus\ker(1-P)$ for the alternating pattern, and for the uniform one whenever $Q_k(1+P)\equiv0\pmod N$ or $N$ is prime; in the exchange branch with even $k$, $Q_k=0$. These conditions are sufficient, not necessary.
\item[(e)] $H_1(\Sigma_{g,k})\cong\Z_N^{4g}\oplus H_1(\Sigma_{0,k})$. For $k=1$, the auxiliary term on the right is given by Example~\ref{ex:rp2}.
\end{enumerate}
\end{theorem}
\begin{proof}
For the one-vertex complex, $\partial_2m=(\rho(\partial\mathfrak r/\partial x)m)_x$ and $\partial_1(m_x)=\sum_x(\rho(x)-1)m_x$. The Fox derivatives of the commutators vanish under $\rho$, and $\partial\mathfrak r/\partial c_j\mapsto P^{2(\eta_1+\dots+\eta_{j-1})}(1+P^{\eta_j})$, giving (b), (e) and $\rho(\mathfrak r)=P^{2\sum\eta_j}$, hence (a). As $P$ is invertible, $\ker d_2=\ker(1+P)$ and $\im d_1=\im(P-1)$, giving Eq.~\eqref{eq:order}; $|\ker(1\pm P)|$ for $\Pd,\Pe$ is computed without dividing by non-units. For (d), coordinates $(m_1,\dots,m_{k-1},\bar m)$ with $\bar m=\sum_j\Lambda_jm_j$ ($\Lambda_j\in\{1,-P^{-1}\}$) put the $\bar m$-component of $\im d_2$ in the form $Q_\eta(1+P)m$, with $Q_\eta=0$ (alternating) or $Q_k$ (uniform); when it vanishes the quotient splits, and for prime $N$ the order fixes the structure.
\end{proof}

\begin{example}[projective plane]\label{ex:rp2}
\noindent\emph{[derivation; not on the preregistered grid]}
For $\Sigma_{0,1}=\mathbb{RP}^2$, set $M=\Z_N^2$ and $\rho(c)=P$. The relation $c^2=1$ requires $P^2=1$. In the same cellular convention as above the complex is
\[
 M\xrightarrow{\;1+P\;}M\xrightarrow{\;P-1\;}M,
\]
so direct calculation from this complex gives
\[
 \begin{aligned}
 H_1(\mathbb{RP}^2;M_\rho)&=\frac{\ker(P-1)}{\im(1+P)},\\
 |H_1|&=\frac{|\ker(P-1)|\,|\ker(1+P)|}{N^2}.
 \end{aligned}
\]
In the diagonal branch consistency is $u^2\equiv1\pmod N$, including $u=\pm1$ for every $N$. With $d_\rho=\gamma_-\gamma_+/N$, a positive integer for these consistent cases,
\[
 H_1\cong\Z_{d_\rho}\oplus\Z_{d_\rho},\qquad |H_1|=d_\rho^2.
\]
In particular, $u=\pm1$ gives $d_\rho=\gcd(2,N)$. In the exchange branch $P^2=-1$, so consistency requires $N=2$; then $H_1=0$ and $|H_1|=1$. These are homology statements, not a construction of a lattice code or new sealed grid data.
\end{example}
\begin{proof}
The one-cell presentation has attaching word $c^2$, whose Fox derivative is $1+c$; hence the two differentials above. Their product is $P^2-1$, so consistency puts $\im(1+P)$ inside $\ker(P-1)$. The order follows by taking the quotient and using $|\im(1+P)|=N^2/|\ker(1+P)|$. For $\Pd$ with $u^2=1$, one has $u^{-1}=u$. In its two cyclic components the kernels of $P-1$ have orders $\gamma_+$ and $\gamma_-$, and the images of $1+P$ have orders $N/\gamma_-$ and $N/\gamma_+$, respectively. Each quotient is cyclic of order $d_\rho$; this also proves its integrality. For $u=\pm1$, $\gamma_-\gamma_+=N\gcd(2,N)$. For $\Pe$ at $N=2$, $P$ swaps the two components and both $\ker(P-1)$ and $\im(1+P)$ are the diagonal subgroup, so their quotient is trivial.
\end{proof}

\begin{figure*}[!t]
\centering\includegraphics[width=0.98\textwidth]{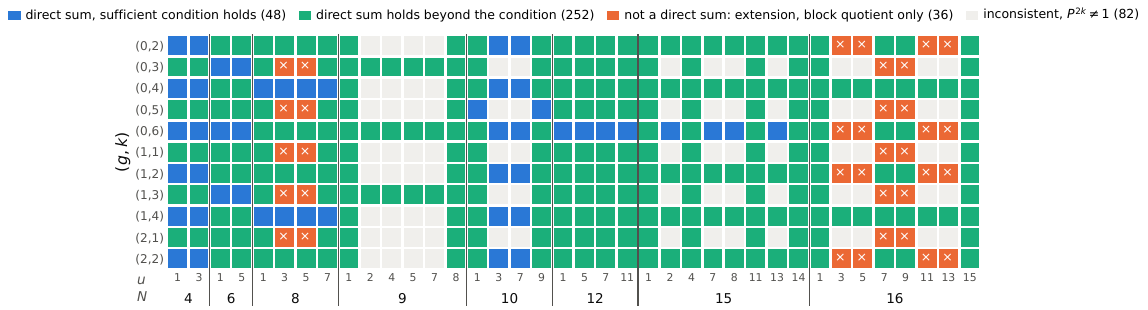}
\caption{Structure of $H_1(\Sigma_{g,k};\Z_N^2{}_\rho)$, uniform pattern, diagonal branch, composite $N$ (columns: units $u$ grouped by $N$; rows: $(g,k)$; sealed E-4 results). Blue: $Q_k(1+P)\equiv0$, direct sum confirmed. Green: the condition fails, the direct sum holds (readings). Orange, $\times$: not a direct sum; the extension of Eq.~\eqref{eq:block} (readings; post-hoc check). Grey: inconsistent, $u^{2k}\not\equiv1$. The alternating pattern, the exchange branch and prime $N$ give direct sums throughout the grid. None of these local systems is realized by a lattice code here.}
\label{fig:h1}
\end{figure*}

\paragraph{Scope.} Lattice codes exist here only for the alternating local system on the Klein bottle, $(g,k)=(0,2)$ (Sec.~\ref{sec:codesK}). For every other surface and local system---handles, $k=1$ or $k\ge3$, and the uniform pattern on the Klein bottle itself, whose holonomy along the rows would be $P^2$---Theorem~\ref{thm:homology} and the separate projective-plane calculation in Example~\ref{ex:rp2} count homology only. A local cross-cap can be realized as a slit whose two lips are glued by a reflection; its ends merge into a vertex of cone angle $4\pi$, where the compatible gluings would have to be classified anew. If such a code exists and its logical module is the twisted homology, its dimension is $|H_1|^{1/2}$, e.g.\ 9 (two qutrits) for $N=3$, $k=4$ in the exchange branch (Table~\ref{tab:H1}); we do not claim that every lattice realization would have this logical module.

\begin{table}[!t]
\caption{Twisted homology of $\Sigma_{g,k}$ (E-4). dim $=|H_1|^{1/2}$, the dimension a code would have if its logical module were $H_1$; a lattice code exists only for the alternating $(0,2)$ row (the Klein bottle of Sec.~\ref{sec:codesK}). $n_k$ as in Remark~\ref{rem:fusion}. $^{*}$Reading: not of direct-sum form.}
\label{tab:H1}
\centering\scriptsize\setlength{\tabcolsep}{2.6pt}
\begin{tabular}{rrllllr}
\toprule
$N$ & $u$ & branch & pattern & $(g,k)$ & $H_1$ & dim\\
\midrule
2 & 1 & exch. & uniform & $(0,4)$ & $\Z_2^6$ & 8\\
3 & 1 & exch. & altern. & $(0,2)$ & $0$ & 1\\
3 & 1 & exch. & uniform & $(0,3)$ & inconsistent & ---\\
3 & 1 & exch. & uniform & $(0,4)$ & $\Z_3^4$ ($n_4=9$) & 9\\
3 & 1 & exch. & uniform & $(0,6)$ & $\Z_3^8$ ($n_6=81$) & 81\\
3 & 1 & exch. & altern. & $(1,2)$ & $\Z_3^4$ ($n_2=1$) & 9\\
4 & 1 & exch. & uniform & $(0,4)$ & $\Z_2^2\oplus\Z_4^4$ & 32\\
5 & 2 & diag. & uniform & $(0,3)$ & inconsistent & ---\\
7 & 2 & diag. & uniform & $(0,3)$ & $\Z_7^2$ & 7\\
8 & 3 & diag. & uniform & $(0,3)$ & $\Z_8^4$\,$^{*}$ & 64\\
\bottomrule
\end{tabular}
\end{table}

\begin{remark}[charge-conjugation defects: consistency only]\label{rem:fusion}
For odd $N$, $C$ fixes only the vacuum, so the counting of symmetry defects \cite{BBCW2019} gives one $C$-defect type $\cD$, with $\cD\times\cD=\sum_{(A,B)\in\Z_N^2}(A,B)$ and quantum dimension $N$ (for defects of other anyon symmetries of $\Z_N$ models see Refs.~\cite{YouWen2012,BarkeshliJianQi2013}); $k$ such defects ($k$ even) have $n_k=N^{k-2}$ vacuum channels, and in the exchange branch with even $k$, $|H_1|=(N^{2g}n_k)^2$ by Eq.~\eqref{eq:order} (432 grid cases). This is one count written twice: consistent with, not a proof of, cross-caps acting as charge-conjugation defects. For even $N$, $C$ fixes the four labels $\{0,N/2\}^2$, so the same counting gives four $C$-defect types, related by fusion with abelian anyons and hence of equal dimension, with $\sum d^2=N^2$, i.e.\ $d=N/2$ (derivation, not computed).
\end{remark}

\subsection{Around a cross-cap: the double-winding monodromy}\label{sec:mono}

\begin{figure}[!t]
\centering\includegraphics[width=\columnwidth]{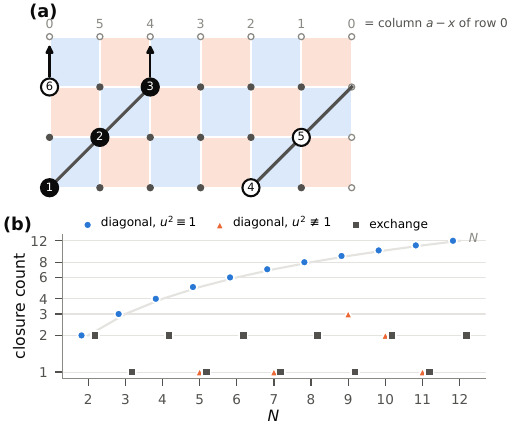}
\caption{(a) $Z$-type double-winding path on $6\times3$, $a=0$ (diagonal branch), sites in order. The first pass (filled) runs from $x_0=0$; from its last site the string enters the seam plaquette through the NW far corner, i.e.\ row 0 at column $a-x=4$ (top labels), where the second pass (open) starts; it re-enters at column $0=x_0$. Here $h=u_{x_0'}u_{x_0}=u^2$. (b) Closure counts of all 1800 registered paths (sealed E-4 results) by $N$ and branch, diagonal-branch paths split by whether $u^2\equiv1$: each group has one value, $\gcd(u^2-1,N)$ (diagonal) or $\gcd(2,N)$ (exchange). Grey line: $N$.}
\label{fig:mono}
\end{figure}

A $Z$-type path from column $x_0$ of row 0 runs through $(x_0+y,y)$, $y<L_y$ (a $Z$ on $(x,y)$ fails to commute only with the plaquettes $(x,y)$ and $(x-1,y-1)$, of colour $x+y$), enters the seam plaquette above its last site and continues, as in the proof of Theorem~\ref{thm:deg}(ii), from its NW far corner at column $x_0'=a-x_0-L_y+1$; a second pass returns it to $x_0$ (Fig.~\ref{fig:mono}a): twice through the seam, once around a cross-cap. $X$-type paths run along $(-1,1)$. The \emph{closure count} is the number of strings on the $2L_y$ path sites, with only $Z$ (or only $X$) powers, that commute with all stabilizers; it equals $|\ker(h-1)|$ for the multiplier $h$ picked up once around.

\begin{proposition}[double-winding monodromy]\label{prop:mono}
For all 540 codes with $N\le12$, all units, $(L_x,L_y)\in\{(4,3),(4,4),(6,3)\}$, $a\le3$, and all 1800 registered paths, the closure count is $\gcd(u^2-1,N)$ in the diagonal and $\gcd(2,N)$ in the exchange branch. The multiplier read generator by generator is $h\in\{u^2,u^{-2}\}$ on all 720 diagonal-branch paths and $h=-1$ on all 1080 exchange-branch paths; for $N>2$, $h=-1$ occurs in the diagonal branch exactly for $u^2\equiv-1$ (64 paths: $N=5$, $u=2,3$; $N=10$, $u=3,7$).
\end{proposition}
The closure counts were pre-registered; $h$ is a reading, equal on all paths to $h=u_{x_0'}u_{x_0}$: the landing columns sum to $a-L_y+1$ and have equal parity in the diagonal branch ($h=u^{\pm2}$) and opposite parity in the exchange branch ($h=u\cdot(-1/u)=-1$).

\smallskip
These readings concern the \emph{external} monodromy of the defined strings: on the registered lattices (derivation and reading), a charge carried once around a cross-cap is acted on by $P^{\pm2}$, the sign depending on the direction---$C$ in the exchange branch, the multiplier $u^{\pm2}$ in the diagonal branch ($C$ if $u^2\equiv-1$). They do not determine the defect type, the cross-cap states, or $F$ and $R$ data (Remark~\ref{rem:fusion}).

\section{Cross-caps and magic: a conditional boundary}\label{sec:magic}

A \emph{stabilizer state} of $n$ qudits is, up to phase, the unique common eigenvector of an abelian group of $N^n$ Weyl operators, i.e.\ of a Lagrangian subgroup $\cG\subset\Z_N^{2n}$ ($\cG^\perp=\cG$) with phases; a \emph{logical stabilizer state} is a code state fixed by a maximal abelian group of logical Weyl operators.

\begin{corollary}[conditional no-go]\label{cor:nogo}
In the operation model $\Op$ every operation maps stabilizer states to stabilizer states, and the outcome statistics of any protocol in $\Op$ applied to a stabilizer input can be computed efficiently classically. Hence, within $\Op$, cross-cap geometry---whatever the seam action $P$, the number of cross-caps and the monodromy---produces no non-stabilizer state from stabilizer inputs, and no protocol in $\Op$ maps a logical stabilizer state of a code of this type to a non-stabilizer state.
\end{corollary}
\begin{proof}
This is the qudit Gottesman--Knill theorem \cite{Gottesman1998,Gottesman1999,Hostens2005,deBeaudrap2013} restricted to $\Op$; we check the steps, including composite $N$. Clifford unitaries, among them qudit permutations and single-site dressings, map Weyl operators to Weyl operators up to phases, hence Lagrangian groups to Lagrangian groups. Measuring $W(w)$ with outcome $\xi$ on the state of $\cG$ gives, if $\xi$ has non-zero probability, a state stabilized by the elements of $\cG$ that commute with $W(w)$ and by $\xi^{-1}W(w)$, i.e.\ by $\cG'=(\cG\cap w^\perp)+\langle w\rangle$. $\cG'$ is isotropic, and since $(\mathcal A\cap\mathcal B)^\perp=\mathcal A^\perp+\mathcal B^\perp$ for subgroups of $\Z_N^{2n}$ and $\langle w\rangle\subset w^\perp$, the modular law gives $\cG'^\perp=(\cG+\langle w\rangle)\cap w^\perp=\cG'$: $\cG'$ is Lagrangian. Feed-forward corrections are Weyl or Clifford operations, and tracking $\cG$ with its phases is efficient \cite{deBeaudrap2013}. A logical stabilizer state is a stabilizer state: the preimage in $\cS^\perp$ of a Lagrangian subgroup of $\cL$ is isotropic of order $|\cS|\,|\cL|^{1/2}=N^n$.
\end{proof}
\begin{remark}[Boundary of the operation model]\label{rem:opboundary}
Corollary~\ref{cor:nogo} applies only to protocols in $\Op$. Coherent path superpositions of mobile charge-conjugation defects and their interferometric measurements, as used in a different construction \cite{ShiHanRaghuVishwanath2026}, are outside the present scope and are not implemented here. Gauging a separately specified charge-conjugation symmetry, when consistent symmetry and defect data permit it \cite{BBCW2019}, changes the theory and is outside the present scope. Reflection-enriched data, fractionalization and cross-cap states \cite{BBCJW2020} are outside the present scope and are not fixed by seam monodromy alone. In particular, the additional symmetry structure supporting the charge measurement considered by Cong, Cheng and Wang \cite{CongChengWang2017} is not constructed from our gluing. External magic states \cite{Veitch2014} are excluded by assumption: protocols supplied with them are outside the present claim. Moving a defect by the Clifford code deformations already allowed in $\Op$ does not by itself supply coherent path control. The label relation $\Pe^2=C$ alone supplies none of these additional structures. These exclusions neither establish impossibility outside $\Op$ nor demonstrate any such resource in the lattice codes studied here.
\end{remark}

The closure argument is the standard stabilizer argument; the contribution here is its geometric setting and boundary. Theorem~\ref{thm:class} classifies commuting gluings within the local monomial far-corner ansatz, and Proposition~\ref{prop:gates} identifies the dressed isometries on the registered grid within its single-site dressing class. For every realization and protocol satisfying $\Op$, the obstruction is uniform in the seam action, the number of cross-caps and the monodromy, because these parameters do not alter stabilizer closure under the permitted operations. This does not assert that arbitrary physical motions or measurements of cross-caps admit an implementation in $\Op$. The argument needs no classification of locality-preserving gates \cite{BravyiKoenig2013} (we neither use nor extend that result to composite $N$ or non-orientable codes).

\subsection{What the post-measurement states show}

For odd $N$ we use the Wigner function of Gross \cite{Gross2006}, defined by $\cW_\rho(q,p)=\mathrm{Tr}[\rho A(q,p)]/N$ and $A(q,p)|j\rangle=\omega^{2p(j-q)}|2q-j\rangle$, so that $A(0,0)=C$. Let $\psi_\pm(j)=(|j\rangle\pm|-j\rangle)/\sqrt2$ ($j\neq0$) and $N_j=N/\gcd(j,N)$; for prime $N$ and $\lambda=u^2\neq1$ of order $r$, let $\phi_{O,\ell}=r^{-1/2}\sum_{m<r}e^{-2\pi i\ell m/r}|\lambda^my\rangle$ be the eigenstates of $\mu_\lambda|x\rangle=|\lambda x\rangle$ on an orbit $O=\{\lambda^my\}\subset\Z_N^\times$.

\begin{proposition}[Wigner negativity]\label{prop:wigner}
(a) The minimum of $\cW$ is $-1/N$ for $\psi_-(j)$, at $(0,0)$, and $-\cos(\pi/N_j)/N$ for $\psi_+(j)$. (b) Measuring $C$ on $|j\rangle$ gives $\psi_\pm(j)$ with probability $1/2$ each; measuring $Z+Z^\dagger$ on $|+\rangle$ gives $|0\rangle$ with probability $1/N$ and $\psi_+(j)$ with probability $2/N$, $j=1,\dots,(N-1)/2$. (c) For prime $N$, $\min\cW(\phi_{O,\ell})\le-1/((N-1)r^2)$, and measuring $\mu_\lambda$ on $|x\rangle$, $x\in O$, gives each eigenvalue with probability $1/r$.
\end{proposition}
\begin{proof}
(a), (b) by direct computation: $\cW$ vanishes off $q\in\{0,\pm j\}$, $\cW(0,p)=\pm\cos(4\pi pj/N)/N$, $\cW(\pm j,p)=1/(2N)$. (c) A subset of $\Z_N$ ($N$ prime) with at least two elements that is closed under midpoints is all of $\Z_N$; as $0\notin O$, some $x\neq x'\in O$ have their midpoint $q\notin O$. In that column $\sum_p\cW(q,p)=\rho_{qq}=0$, $\sum_p\cW(q,p)^2=m_q/(Nr^2)$ with $m_q=|O\cap(2q-O)|\ge2$, and $|\cW|\le1/N$, so the at most $N-1$ negative entries give $\min\cW\le-m_q/(2(N-1)r^2)$.
\end{proof}

\begin{figure}[!t]
\centering\includegraphics[width=\columnwidth]{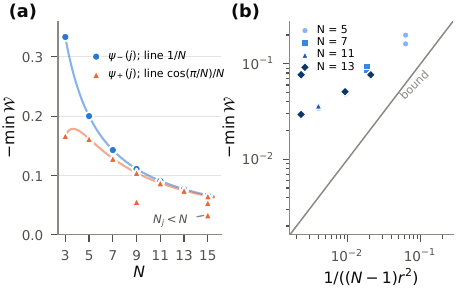}
\caption{Wigner negativity of post-measurement states (sealed E-4 results; $N$ odd). (a) $-\min\cW$ of $\psi_\pm(j)$ for the 28 registered $(N,j)$, against $1/N$ and $\cos(\pi/N)/N$; the triangles below the orange line are the four cases with $N_j<N$ (two coincide at $N=15$). (b) Multiplier eigenstates for prime $N$ (116 states) against the bound $1/((N-1)r^2)$: all above the line, by a factor $\ge2.59$.}
\label{fig:wigner}
\end{figure}

\smallskip
A pure state of odd dimension has a non-negative Wigner function iff it is a stabilizer state \cite{Gross2006}, and negativity is necessary for magic-state distillation to help \cite{Veitch2012}; $\psi_-(1)$ at $N=3$ is the strange state, $\min\cW=-1/3$ \cite{Veitch2014}. Proposition~\ref{prop:wigner} therefore certifies only that one non-stabilizer measurement---of $C$, or of a charge that does not distinguish a label from its conjugate, such as $Z+Z^\dagger$---yields non-stabilizer states with fixed probabilities. That measurement lies outside $\Op$. Nothing here shows distillability, fault tolerance, or a universal gate set.

\subsection{Symmetry protection, by hand}\label{sec:sym}

Could the orientation reversal itself protect a measurement that does not distinguish a charge from its conjugate? Take a tunnelling term $H_t=-(tV+t^*V^\dagger)$. A symmetry of the Hamiltonian can pin $t$ in three ways. (i) A \emph{unitary} $C$ with $CVC^{-1}=V^\dagger$ requires $t=t^*$. (ii) An \emph{antiunitary} $T$ also conjugates scalars: if $TVT^{-1}=V^\dagger$, $H_t$ is invariant for every complex $t$; only if $V$ is $T$-even, $TVT^{-1}=e^{i\vartheta}V$, does invariance give $t=e^{i\vartheta}t^*$. This is the mechanism of Cong, Cheng and Wang: in their fractional quantum spin Hall realization ``the time-reversal symmetry exchanges the two layers'', the charge $e$ is time-reversal invariant, and so ``the tunneling amplitude of $e$ has to be real''; ``the requirement that $t$ is real is beyond topological protection'' \cite{CongChengWang2017}. (iii) A \emph{unitary spatial reflection} that maps the coupling path to itself and $V$ to $V^\dagger$ pins $t$ exactly as $C$ does: on the literal gluing $U_\sigma F_pU_\sigma^\dagger=F_{\sigma(p)}^\dagger$ (Proposition~\ref{prop:sigma}), so a $\sigma$-invariant term $-(tF_p+t^*F_p^\dagger)$ on a $\sigma$-fixed plaquette ($y=(L_y-1)/2$ and $2x\equiv a-1\pmod{L_x}$, e.g.\ $x=0,2$ at $4\times3$, $a=1$) has $t$ real.

The Klein-bottle gluing, by contrast, is a gluing rule, not a symmetry of the Hamiltonian, and by itself constrains no local amplitude. The three pinning symmetries are microscopic symmetries that must be imposed; they are equally available on orientable lattices (a mirror through the coupling on a plane or torus) and fragile to disorder. If a Klein-bottle lattice carries such a reflection, its protection is symmetry protection, not topological, and does not come from non-orientability. Within $\Op$ the answer is negative; Corollary~\ref{cor:nogo} and $\Op$ are unchanged. This is a hand assessment; the exclusions of $\Op$ apply, among them reflection-enriched data \cite{BBCJW2020} and path superpositions of mobile defects \cite{ShiHanRaghuVishwanath2026}.

\section{Relation to neighbouring work}\label{sec:lit}

\paragraph{Homological codes.} Lei shows that a qudit homological code on any 2-complex, for any qudit dimension $D$, has dimension $|H_1(\Sigma;\Z_D)|$ \cite{Lei2023qudit}; our edge-register codes are instances. Chan, Teo and Ryu relate the degeneracy on non-orientable surfaces to parity-invariant anyons \cite{ChanTeoRyu2016}; Theorem~\ref{thm:deg} establishes this count, $|\ker(1-P)|$, for the seam actions of local lattice gluings, including the order-four ones. Bombin and Martin-Delgado state that ``for nonorientable surfaces it is impossible to construct homological codes based on qudits of dimension $D>2$'' \cite{BombinMartinDelgado2007}, tracing this to ``the torsion subgroup'' $\Z_2$ of $H_1$ through their requirement that the stabilizer subspace have ``a basis which is a linearly independent set'' in $\Z_D^{2n}$, so that the code is an integer number of $D$-dimensional qudits (their example: $D=4$ on the projective plane). For odd $D$ that torsion is absent ($H_1(K;\Z_D)=\Z_D$). Their no-go concerns their construction, a free linear subspace of $\Z_D^{2n}$. Codes whose stabilizer is a module, which allows $\Z_N$ torsion, exist on non-orientable surfaces for every $D$: Lei's theorem gives them in general, and explicitly the edge register on $K$, the toric code of a cellulation, has commuting stabilizers and, for odd $N$, dimension $N$. What fails is therefore the broad reading that no non-orientable code with $D>2$ exists; we do not claim that the two constructions coincide. For even $D$ the obstruction concerns only the integer-number-of-qudits requirement, which the $\Z_{\gcd(2,D)}$ torsion violates; our edge-register codes (dimension $2N$) do not impose it and are the twisted homology of Proposition~\ref{prop:L} with $P=\mathrm{diag}(1,-1)$. The site-register $u$-codes are not cellular codes in the usual sense (their checks change type and carry unit multipliers across the seam), but on the grid their logical modules are twisted homology groups as well.

\paragraph{Cross-cap gates and reflection-enriched order.} Cross-cap defects implement a fault-tolerant logical Hadamard in the $\Z_2$ surface code \cite{Kobayashi2024crosscap}; Proposition~\ref{prop:gates} is a $\Z_N$ comparison for lattice isometries. Barkeshli, Bonderson, Cheng, Jian and Walker develop reflection- and time-reversal-enriched phases on non-orientable manifolds, with cross-cap states and fractionalization data \cite{BBCJW2020}. Their Sec.~II acknowledges order-four and higher reflection actions before adopting an order-two convention for the subsequent analysis. Our exchange-branch $P^2=C$ is an explicit local-system monodromy of the stated lattice gluing; we do not claim order-four reflection itself as new or identify our cross-caps with their defect classes.

\paragraph{Quotient codes and twisted compactification.} Hsin~\cite{Hsin2026quotient} formulates monomial stabilizer-code quotients through twisted compactification, with a transfer theorem for logical operators and distance, a Clifford restriction for descended monomial symmetries, and a separate discussion of composite dimensions. Our focus is the explicit local far-corner seam ansatz of the $\Z_N$ plaquette model: its commuting $u$-family, phase-sensitive consistency, and the noncommuting literal seam with bulk-locked clock spectra. These are model-resolved questions, rather than a general quotient transfer theorem. The non-Clifford constructions of Ref.~\cite{Hsin2026quotient} use additional cup-product circuits or fermionic pumps outside $\Op$; they do not contradict the stabilizer-closure statement of Corollary~\ref{cor:nogo}.

\paragraph{Clifford limits and routes to magic.} Webster and Bartlett prove non-universality for combinations of defect braiding and locality-preserving logical operators, including Abelian quantum-double models; their stronger Clifford restriction applies to a specified standard defect encoding \cite{WebsterBartlett2020}. Corollary~\ref{cor:nogo} instead follows directly from the explicitly chosen stabilizer operation model. In the charge-$4e$ superconductor of Shi, Han, Raghu and Vishwanath the $hc/4e$ fluxes are charge-conjugation defects of a chiral $\Z_3$ order: braiding gives the many-qutrit Clifford group, and a single-probe interferometric measurement prepares magic states \cite{ShiHanRaghuVishwanath2026}. Our exchange-branch cross-caps also have double-winding monodromy $C$, but within $\Op$ a cross-cap moves only by code deformation, and no coherent superposition of paths arises.

\paragraph{Other neighbours.} Lattice twists on a Klein bottle must represent its fundamental group \cite{SeibergZhang2026}. Takahashi, Li and Yoshida construct an order-two locality-preserving Clifford realization of electric--magnetic duality for odd $N$ and exclude it for even $N$ on periodic toric-code lattices sufficiently large compared with the locality range; their obstruction locally emulates a cross-cap \cite{TakahashiLiYoshida2026}; Proposition~\ref{prop:gates} asks instead which isometries of the Klein-bottle codes admit single-site dressings; its condition for a non-trivial colour exchange, $u^2\equiv-1$, comes from compatibility with the seam action. Self-dual chains appear in parafermion chains \cite{LiYangTuCheng2015}, twist-defect chains \cite{YuChenRoyTeo2017} and edges of Wen's model \cite{HoCincioMoradiGaiottoVidal2015}; lattice defects implement Kramers--Wannier duality in the Ising model \cite{AasenMongFendley2016}, and lattice reflections implement it in one dimension \cite{PaceDelfinoLamAksoy2025}; specific here is its realization by the Klein-bottle reflection $\sigma$ exchanging the two plaquette rows of an obstructed gluing.

\section{Discussion}\label{sec:discussion}

The classification covers local, monomial, far-corner gluings only; non-local gluings, multi-site far corners, modified bulk plaquettes and the glide gluings of Ref.~\cite{Wang2026paper3} at $N>2$ were not studied. Grid statements are not extrapolated (the odd-$N$ edge distance $d_X=L_x$ fails beyond the grid); the site-register logical module is verified on the grid and only argued in general; the gate results hold within the single-site dressing class; lattice codes exist only for the alternating Klein bottle, and elsewhere we only count homology; the ring degeneracy is not claimed to be protected. The no-go holds within $\Op$, and the symmetry assessment is by hand. Producing a non-stabilizer resource from stabilizer inputs requires an operation outside $\Op$; the extensions listed in Remark~\ref{rem:opboundary} are outside the present scope, and their availability or sufficiency in these lattice codes is not established here. Thresholds, decoders and noise were not considered.

\section*{Acknowledgments}
The research program was designed and directed by the author. Derivations, pre-registered predictions, code, figures and drafting were produced with AI assistance (Anthropic Claude). Independent AI instances, including OpenAI Codex, assisted with cross-checks; these were AI-assisted checks, not independent human verification or journal peer review. The reports and dispositions are archived with the accompanying material (Appendix~\ref{app:repro}). The author takes full responsibility for the content. This work received no external funding.

\appendix

\section{Evidence, provenance and reproducibility}\label{app:repro}

\subsection{Registered grids}\label{app:grids}
E-1: $N\in\{2,\dots,13,16,17,25,26\}$, $L_x\in\{3,\dots,8\}$, $L_y\in\{3,4,5\}$, $a\le3$, all units. E-2: $N\in\{2,\dots,10,12\}$, $L_x\in\{3,\dots,8\}$, $L_y\in\{2,\dots,5\}$, $a<\min(L_x,4)$; 920 cases. E-3: site register $N\in\{2,\dots,12,15,16\}$, all $u$, $L_x\in\{4,6\}$, $L_y\in\{3,4,5\}$, $a\le3$, plus $8\times3$, $a\le1$ (1586 codes); edge register $N\le12$, $L_x\in\{3,\dots,6\}$, $L_y\in\{3,4,5\}$, $a<\min(L_x,4)$ (495 codes); isometries on 1378 site ($N\le12$ and $16$) and 264 edge codes ($L_x\in\{4,6\}$); distances on 135 site and 30 edge codes. E-4: Sec.~\ref{sec:cross}, Propositions~\ref{prop:mono}, \ref{prop:wigner}.

\subsection{Pre-registration and runs}\label{app:prereg}
E-0 is a calibration without pre-registration. For E-1 to E-4 the predictions, anchors, grids and falsification criteria were frozen in \texttt{prereg/eK\_prereg.md} before the stage's runs in this package (sha256 and time in \texttt{logs/eK\_prereg\_stamp.txt}, Table~\ref{tab:seals}; records, not cryptographic proofs of priority) and were not changed afterwards. Each lists what was known beforehand, which is not counted as prediction: for E-1, E-0 readings on the Fourier-dressed gluing (code dimensions at $4\times3$, $4\times4$, $N\le6$, part of H3; the $N=2$ odd-$L_x$ sign, part of H4); for E-2, the literal-gluing anatomy and the $N=2$ results; for E-3, the E-1 results and the $N=2$ codes; for E-4, the seam action $P$, planning-stage checks by another AI instance outside the package ($|H_1|=N^{2(k-2)}$ for $k$ identical exchange-branch cross-caps on the sphere, $N=3,5,7,9,15$, even $k\le6$; the $N=3$ Wigner minima $-1/6$, $-1/3$) and a hand computation of the uniform $N=8$, $u=3$, $k=2$ case, which overlap H1(b) and H4(a). The seam action $P$, $\Pe^2=C$ and the phase statement of Theorem~\ref{thm:deg}(iii) were not pre-registered (E-1 registered $|\cR|$ and the phase-free frustration rule); $P$ first appears in a pre-registration as the stated basis of E-3 H1.

Each sealed run is one execution of \texttt{code/run\_eK.sh} that started with no result files of its stage (the manifest lists the files present at the start). It was preceded by full development or rehearsal runs, whose results were inspected and in part reviewed, and code, gates and check counts changed between them. For example, E-0 ran twice against the v0.4 targets of Ref.~\cite{Wang2026paper3} on 2026-09-30 and again on 2026-10-01; E-1 ran on 2026-09-30 at 16:37 UTC (57\,018 checks; reviewed) and 19:21 (59\,042); E-2 was sealed, reviewed and rerun with a changed gate and identical checks (superseded seal \texttt{6d0c3e1f\dots}); E-3 had two rehearsals (the reviewed one with 36\,329 checks; sealed run 34\,762); after E-4's reviewed rehearsal, 1800 cross-checks of $h$ were added.

\subsection{Evidence by result}\label{app:evidence}
The proofs of draft v0.1 were re-derived by an independent AI reviewer; those added in v0.2 (Corollary~\ref{cor:nogo}, the edge-register part of Proposition~\ref{prop:L}) were checked by the external AI review of draft v0.2 but have not been re-derived independently by a human. Sealed checks per hypothesis (correlated within a case): E-1 H1--H4 4182, 261, 4536, 9252; E-2 H1--H5 3105, 7600, 196, 399, 6040; E-3 H1--H5 7531, 11\,566, 390, 4926, 4602; E-4 H1--H4 5720, 864, 5400, 414.
\emph{Theorem~\ref{thm:class}:} complete enumeration of all far-corner assignments in 16 small cases ($644\,782\,854$ gluings: $N=2$, $L_x\le6$; $N=3$, $L_x\le4$; $N=4,5$, $L_x=3$; $L_y=3$, $a\le1$) finds exactly the $u$-family; a pruned exact search returns the predicted sets on all 1152 grid cases; for $N\le30$ ($L_x=3,5$) odd-$L_x$ gluings exist exactly for $N\in\{2,5,10,13,17,25,26,29\}$.
\emph{Theorem~\ref{thm:deg}:} 7272 stabilizer-group orders agree between three algorithms (Howell form \cite{Howell1986}, integer determinant, Smith form); state-vector ranks equal the algebraic dimension in 21 cases up to $16\,777\,216$ states. Part (ii) rests on 17\,208 sealed property checks of the seam map (colour exchange iff $a+L_y$ even; entries multiplying to $-1$; square $C$ exactly in the exchange branch and for $u^2\equiv-1$; $|\cR|$ fixed labels); the equality of the 239 stored matrices with Eq.~\eqref{eq:P} was checked post hoc; the overall sign, a convention, is not tested. The phase statement of (iii) is a cross-check added after the internal review (846 cases).
\emph{Theorems~\ref{thm:anatomy}, \ref{thm:lock}, Proposition~\ref{prop:sigma}:} 920 grid cases, including the dimensions of $186\,712$ joint eigenspaces ($54\,613$ allowed); complete spectra on nine dense lattices (up to 6561 states) and block-resolved at $N=3$, $3\times3$; Lanczos ground energies and low levels on five lattices (up to $1\,953\,125$ states), ground energies only on three (up to $16\,777\,216$); all 34 ground energies agree with Eq.~\eqref{eq:E0} to $9.5\times10^{-14}$, which also reproduces the six energies and twenty degeneracies of Ref.~\cite{Wang2026paper3}. The ring identity was checked on 68 ring sizes ($N\le8$). Two registered tests were not run (low levels of the three largest lattices; complete ring spectra beyond $N^m=2048$), so (c) is confirmed with reduced coverage.
\emph{Propositions~\ref{prop:L}, \ref{prop:gates}:} invariant factors of all 1586 site and 495 edge codes by two routes (Smith form of the commutation Gram matrix; Howell counting), with $|\cL|$ the square of the dimension of Theorem~\ref{thm:deg}; gates exhaustive on 1378 site and 264 edge codes. For odd $N$, (b) and (c) of Proposition~\ref{prop:gates} follow from (a), so its informative tests are at even $N$; at $N=2$ the groups of Ref.~\cite{Wang2026paper3} are reproduced on its 30 codes.
\emph{Theorem~\ref{thm:homology}:} 2074 cases, 1570 consistent, invariant factors by two routes; (d) holds on its 1282 cases (550 with $N$ prime, implied by the order). The 288 consistent uniform cases outside its conditions (all diagonal, composite $N$) are readings: 252 of direct-sum form, 36 not ($N=8$: 10; $N=16$: 26), e.g.\ $\Z_8^4$ at $N=8$, $u=3$, $(0,3)$, not $\Z_2^2\oplus\Z_4^2\oplus\Z_8^2$; a post-hoc recomputation from Eq.~\eqref{eq:block} reproduces all 288. The alternating $(0,2)$ case reproduces the modules of all E-3 codes; H2 follows from H1(b).
\emph{Propositions~\ref{prop:mono}, \ref{prop:wigner}:} the closure counts separate the branches on 1120 paths; $h$, added after the internal review because closure counts cannot tell $h=-1$ from $u^2$ when $\gcd(2,N)=\gcd(u^2-1,N)$, is a reading; torus control: 28 checks. Wigner: the least negative orbit state is $N=13$, $\lambda=4$ ($-0.0295$ against the bound $-0.0023$); stabilizer states have $\cW\ge-7.4\times10^{-17}$.
\emph{Post hoc} (\texttt{paper/verify\_v02.py}, exact arithmetic, unsealed): Lemma~\ref{lem:sign} ($N=2$, $L_x\le12$, $L_y\le6$, all $a$; $\prod_pF_p^{N/2}$ for $N\le26$), the dressed glide and the exponents of Theorem~\ref{thm:class}, Theorem~\ref{thm:lock}(a) in both orders, the CSS form used for Proposition~\ref{prop:L}, and the $\sigma$-fixed plaquettes of Sec.~\ref{sec:sym}.

\subsection{Engine, seals and gates}\label{app:seals}
A NumPy-only engine represents operators as Weyl strings with exact phases and uses the Howell form \cite{Howell1986} for orders, relations and membership, and dense diagonalization and Lanczos for spectra. E-0 calibrated it at $N=2$ against the six result files of Ref.~\cite{Wang2026paper3} (v0.7, unchanged in v0.8): $8005$ comparisons ($606$ stored as truth values), $1934$ definitional items, $453$ internal cross-checks, 12 self-tests; errors specific to $N>2$ are addressed only by each stage's self-tests and anchors.

\begin{table}[!t]
\caption{Pre-registration freeze, seal (first 16 hex digits; full values in \texttt{SEALS.txt}) and sealing time of each stage (UTC; all dates in 2026); checks are prediction/anchor/cross-check items of the sealed run.}
\label{tab:seals}
\centering\scriptsize\setlength{\tabcolsep}{2.2pt}
\begin{tabular}{llll}
\toprule
stage & prereg frozen & seal & sealed \\
\midrule
E-0 & (calibration) & \texttt{0901726bc670b9df} & 10-01 10:11:53\\
E-1 & 09-30 15:20:37 & \texttt{55136031bc669cbb} & 10-01 09:52:44\\
E-2 & 10-01 14:49:29 & \texttt{559bdcb43a1cdd7d} & 10-02 07:41:10\\
E-3 & 10-02 09:04:58 & \texttt{cf526247f4931553} & 10-02 11:10:21\\
E-4 & 10-02 17:45:44 & \texttt{97811bc292074cfe} & 10-02 18:35:22\\
\midrule
\multicolumn{4}{l}{checks: E-1 21\,108/10\,027/27\,907; E-2 15\,956/13/1\,410;}\\
\multicolumn{4}{l}{E-3 25\,656/2\,408/6\,698; E-4 8\,798/2\,121/7\,330}\\
\bottomrule
\end{tabular}
\end{table}

Each sealed run writes the sha256 of every code file (gate and \texttt{seal.py} included), input and result to \texttt{results/eK\_files.sha256}, whose own sha256 is the seal recorded in \texttt{SEALS.txt} after the gate passed (Table~\ref{tab:seals}); both can be checked with \texttt{sha256sum} alone. The gate \texttt{code/eK\_gate.py} re-evaluates every stored comparison, fixes the number and kinds of checks and, for E-2 to E-4, the coverage case by case; the E-3 and E-4 gates recompute $5196$ and $13\,787$ expected values from the pre-registered formulas. The gates guard against accidents, not against a writer who rewrites results and seal list consistently; mutation tests (unsealed record) behaved as expected. The gate summaries \texttt{results/eK\_gate.json} are not sealed, and running a gate rewrites them. The only complete check is a rerun.

\paragraph{Deviations.} Besides the two E-2 tests not run (Appendix~\ref{app:evidence}), the E-2 ring range $N\le8$ was fixed in code, not in the pre-registration, and 110 of the 7988 off-grid E-1 cross-checks duplicate a repeated random draw (no verdict affected). Grid-only statements, readings and post-hoc checks are labelled where they occur.

\paragraph{Reviews.} Before sealing, an AI instance without the authoring context reviewed each of E-1 to E-4 adversarially, re-implementing the central computations and attacking the gates; its findings (gate defects, two rated P0 in E-1; coverage gaps; an off-grid distance failure; uninformative checks; the blindness of closure counts to $h$) were fixed before sealing or recorded. OpenAI Codex reviewed every stage. Its first E-0 review found two P0 problems---the gate trusted cached result files, and the odd-$L_x$ dressed gluing of Ref.~\cite{Wang2026paper3} is sign-frustrated rather than a code with $\GSD=2$---which led to the seal mechanism, the reruns of E-0 and E-1, and the erratum v0.7 of Ref.~\cite{Wang2026paper3}; on the current seals it found no P0. The external review of draft v0.2 identified a P1 error in the projective-plane domain of Theorem~\ref{thm:homology}; the v0.3 disposition incorrectly summarized that review as having no P1. Draft v0.4 restores $2g+k\ge2$ and separates Example~\ref{ex:rp2}, without changing the sealed results. The correction and the preserved v0.3 files are documented in \texttt{ERRATA\_v0.4.md}; the earlier wording fixes remain in place. For E-2 its point on case-by-case gate coverage led to a changed gate and a full rerun; its other P1 points were adopted. Reports and dispositions are under \texttt{reviews/}, \texttt{E\_campaign/review/} and, for this paper, \texttt{paper/review\_v0.1/} and \texttt{paper/review\_codex\_v0.2/}.

\paragraph{Data availability.} The accompanying ancillary archive, \texttt{paperE\_reproducibility\_v0.6.zip}, contains the sealed E-0--E-4 code, inputs, results and supporting records in their campaign-relative layout, together with the number-tracing and post-hoc verification scripts. Historical reports retain their original statements; the scientific scope and corrections in this manuscript take precedence. Version DOI for the full reproducibility archive: \url{https://doi.org/10.5281/zenodo.24185575}.

\paragraph{Reproduction.} \texttt{bash code/run\_eK.sh} reruns a stage (sealed wall times 66, 61, 59, 10\,min and 7\,s for E-0 to E-4 on a two-core machine; Python 3.11.15, NumPy 2.4.4; about 4\,GB for E-1, E-2) and gives a new seal; rerunning E-3 invalidates the E-4 seal. \texttt{paper/claims\_numbers.md} traces every number, \texttt{paper/check\_numbers.py} prints them from the sealed files after verifying the seals (marking unsealed inputs), \texttt{paper/make\_figs.py} draws Figs.~\ref{fig:h1}--\ref{fig:wigner}, and \texttt{paper/verify\_v02.py} holds the post-hoc checks. No second human has checked derivations or computations.

\bibliographystyle{unsrtnat-init}
\bibliography{references}

\begin{thebibliography}{34}
\providecommand{\natexlab}[1]{#1}
\providecommand{\url}[1]{\texttt{#1}}
\expandafter\ifx\csname urlstyle\endcsname\relax
  \providecommand{\doi}[1]{doi: #1}\else
  \providecommand{\doi}{doi: \begingroup \urlstyle{rm}\Url}\fi

\bibitem[Wang(2026)]{Wang2026paper3}
J.~Wang.
\newblock An exactly solvable obstructed seam on a lattice {K}lein bottle:
  bulk-locked {M}ajorana rings and register-dependent toric codes, 2026.
\newblock URL \url{https://arxiv.org/abs/2610.35931}.
\newblock Version 1.

\bibitem[Wen(2003)]{Wen2003plaquette}
X.-G. Wen.
\newblock Quantum orders in an exact soluble model.
\newblock \emph{Phys. Rev. Lett.}, 90:\penalty0 016803, 2003.
\newblock \doi{10.1103/PhysRevLett.90.016803}.

\bibitem[Lei(2023)]{Lei2023qudit}
Z.~Lei.
\newblock Qudit surface codes and hypermap codes.
\newblock \emph{Quantum Inf. Process.}, 22:\penalty0 297, 2023.
\newblock \doi{10.1007/s11128-023-04060-8}.

\bibitem[Chan et~al.(2016)Chan, Teo, and Ryu]{ChanTeoRyu2016}
A.~P.~O. Chan, J.~C.~Y. Teo, and S.~Ryu.
\newblock Topological phases on non-orientable surfaces: twisting by parity
  symmetry.
\newblock \emph{New J. Phys.}, 18:\penalty0 035005, 2016.
\newblock \doi{10.1088/1367-2630/18/3/035005}.

\bibitem[Barkeshli et~al.(2020)Barkeshli, Bonderson, Cheng, Jian, and
  Walker]{BBCJW2020}
M.~Barkeshli, P.~Bonderson, M.~Cheng, C.-M. Jian, and K.~Walker.
\newblock Reflection and time reversal symmetry enriched topological phases of
  matter: path integrals, non-orientable manifolds, and anomalies.
\newblock \emph{Commun. Math. Phys.}, 374:\penalty0 1021--1124, 2020.
\newblock \doi{10.1007/s00220-019-03475-8}.

\bibitem[Kobayashi and Zhu(2024)]{Kobayashi2024crosscap}
R.~Kobayashi and G.~Zhu.
\newblock Cross-cap defects and fault-tolerant logical gates in the surface
  code and the honeycomb {F}loquet code.
\newblock \emph{PRX Quantum}, 5:\penalty0 020360, 2024.
\newblock \doi{10.1103/PRXQuantum.5.020360}.

\bibitem[Hsin(2026)]{Hsin2026quotient}
P.-S. Hsin.
\newblock Twisted compactification, quotient quantum codes and logical gates,
  2026.
\newblock URL \url{https://arxiv.org/abs/2611.09538}.
\newblock Version 1.

\bibitem[Cong et~al.(2017)Cong, Cheng, and Wang]{CongChengWang2017}
I.~Cong, M.~Cheng, and Z.~Wang.
\newblock Universal quantum computation with gapped boundaries.
\newblock \emph{Phys. Rev. Lett.}, 119:\penalty0 170504, 2017.
\newblock \doi{10.1103/PhysRevLett.119.170504}.

\bibitem[Shi et~al.(2026)Shi, Han, Raghu, and
  Vishwanath]{ShiHanRaghuVishwanath2026}
Z.~D. Shi, Z.~Han, S.~Raghu, and A.~Vishwanath.
\newblock Charge-$4e$ superconductor with parafermionic vortices: A path to
  universal topological quantum computation, 2026.
\newblock URL \url{https://arxiv.org/abs/2602.06963v2}.
\newblock Version 2.

\bibitem[You and Wen(2012)]{YouWen2012}
Y.-Z. You and X.-G. Wen.
\newblock Projective non-{A}belian statistics of dislocation defects in a
  {$\mathbb{Z}_N$} rotor model.
\newblock \emph{Phys. Rev. B}, 86:\penalty0 161107(R), 2012.
\newblock \doi{10.1103/PhysRevB.86.161107}.

\bibitem[Kitaev(2003)]{Kitaev2003}
A.~Y. Kitaev.
\newblock Fault-tolerant quantum computation by anyons.
\newblock \emph{Ann. Phys.}, 303:\penalty0 2, 2003.
\newblock \doi{10.1016/S0003-4916(02)00018-0}.

\bibitem[Delmastro and Gomis(2021)]{DelmastroGomis2021}
D.~Delmastro and J.~Gomis.
\newblock Symmetries of {A}belian {C}hern--{S}imons theories and arithmetic.
\newblock \emph{JHEP}, 03:\penalty0 006, 2021.
\newblock \doi{10.1007/JHEP03(2021)006}.

\bibitem[Li et~al.(2015)Li, Yang, Tu, and Cheng]{LiYangTuCheng2015}
W.~Li, S.~Yang, H.-H. Tu, and M.~Cheng.
\newblock Criticality in translation-invariant parafermion chains.
\newblock \emph{Phys. Rev. B}, 91:\penalty0 115133, 2015.
\newblock \doi{10.1103/PhysRevB.91.115133}.

\bibitem[Sun et~al.(2019)Sun, Vekua, Cobanera, and
  Ortiz]{SunVekuaCobaneraOrtiz2019}
G.~Sun, T.~Vekua, E.~Cobanera, and G.~Ortiz.
\newblock Phase transitions in the {$\mathbb{Z}_p$} and {U}(1) clock models.
\newblock \emph{Phys. Rev. B}, 100:\penalty0 094428, 2019.
\newblock \doi{10.1103/PhysRevB.100.094428}.

\bibitem[Beverland et~al.(2016)Beverland, Buerschaper, Koenig, Pastawski,
  Preskill, and Sijher]{Beverland2016}
M.~E. Beverland, O.~Buerschaper, R.~Koenig, F.~Pastawski, J.~Preskill, and
  S.~Sijher.
\newblock Protected gates for topological quantum field theories.
\newblock \emph{J. Math. Phys.}, 57:\penalty0 022201, 2016.
\newblock \doi{10.1063/1.4939783}.

\bibitem[Bravyi and K{\"o}nig(2013)]{BravyiKoenig2013}
S.~Bravyi and R.~K{\"o}nig.
\newblock Classification of topologically protected gates for local stabilizer
  codes.
\newblock \emph{Phys. Rev. Lett.}, 110:\penalty0 170503, 2013.
\newblock \doi{10.1103/PhysRevLett.110.170503}.

\bibitem[Barkeshli et~al.(2019)Barkeshli, Bonderson, Cheng, and Wang]{BBCW2019}
M.~Barkeshli, P.~Bonderson, M.~Cheng, and Z.~Wang.
\newblock Symmetry fractionalization, defects, and gauging of topological
  phases.
\newblock \emph{Phys. Rev. B}, 100:\penalty0 115147, 2019.
\newblock \doi{10.1103/PhysRevB.100.115147}.

\bibitem[Barkeshli et~al.(2013)Barkeshli, Jian, and Qi]{BarkeshliJianQi2013}
M.~Barkeshli, C.-M. Jian, and X.-L. Qi.
\newblock Twist defects and projective non-{A}belian braiding statistics.
\newblock \emph{Phys. Rev. B}, 87:\penalty0 045130, 2013.
\newblock \doi{10.1103/PhysRevB.87.045130}.

\bibitem[Gottesman(1998)]{Gottesman1998}
D.~Gottesman.
\newblock The {H}eisenberg representation of quantum computers.
\newblock \emph{arXiv:quant-ph/9807006}, 1998.

\bibitem[Gottesman(1999)]{Gottesman1999}
D.~Gottesman.
\newblock Fault-tolerant quantum computation with higher-dimensional systems.
\newblock \emph{Chaos Solitons Fractals}, 10:\penalty0 1749, 1999.
\newblock \doi{10.1016/S0960-0779(98)00218-5}.

\bibitem[Hostens et~al.(2005)Hostens, Dehaene, and De~Moor]{Hostens2005}
E.~Hostens, J.~Dehaene, and B.~De~Moor.
\newblock Stabilizer states and {C}lifford operations for systems of arbitrary
  dimensions and modular arithmetic.
\newblock \emph{Phys. Rev. A}, 71:\penalty0 042315, 2005.
\newblock \doi{10.1103/PhysRevA.71.042315}.

\bibitem[de~Beaudrap(2013)]{deBeaudrap2013}
N.~de~Beaudrap.
\newblock A linearized stabilizer formalism for systems of finite dimension.
\newblock \emph{Quantum Inf. Comput.}, 13:\penalty0 73, 2013.
\newblock \doi{10.26421/QIC13.1-2-6}.

\bibitem[Veitch et~al.(2014)Veitch, Mousavian, Gottesman, and
  Emerson]{Veitch2014}
V.~Veitch, S.~A.~H. Mousavian, D.~Gottesman, and J.~Emerson.
\newblock The resource theory of stabilizer quantum computation.
\newblock \emph{New J. Phys.}, 16:\penalty0 013009, 2014.
\newblock \doi{10.1088/1367-2630/16/1/013009}.

\bibitem[Gross(2006)]{Gross2006}
D.~Gross.
\newblock Hudson's theorem for finite-dimensional quantum systems.
\newblock \emph{J. Math. Phys.}, 47:\penalty0 122107, 2006.
\newblock \doi{10.1063/1.2393152}.

\bibitem[Veitch et~al.(2012)Veitch, Ferrie, Gross, and Emerson]{Veitch2012}
V.~Veitch, C.~Ferrie, D.~Gross, and J.~Emerson.
\newblock Negative quasi-probability as a resource for quantum computation.
\newblock \emph{New J. Phys.}, 14:\penalty0 113011, 2012.
\newblock \doi{10.1088/1367-2630/14/11/113011}.

\bibitem[Bombin and Martin-Delgado(2007)]{BombinMartinDelgado2007}
H.~Bombin and M.~A. Martin-Delgado.
\newblock Homological error correction: classical and quantum codes.
\newblock \emph{J. Math. Phys.}, 48:\penalty0 052105, 2007.
\newblock \doi{10.1063/1.2731356}.

\bibitem[Webster and Bartlett(2020)]{WebsterBartlett2020}
P.~Webster and S.~D. Bartlett.
\newblock Fault-tolerant quantum gates with defects in topological stabilizer
  codes.
\newblock \emph{Phys. Rev. A}, 102:\penalty0 022403, 2020.
\newblock \doi{10.1103/PhysRevA.102.022403}.

\bibitem[Seiberg and Zhang(2026)]{SeibergZhang2026}
N.~Seiberg and W.~Zhang.
\newblock Tori, {K}lein bottles, and modulo 8 parity/time-reversal anomalies of
  2+1d staggered fermions.
\newblock \emph{JHEP}, 05:\penalty0 264, 2026.
\newblock \doi{10.1007/JHEP05(2026)264}.

\bibitem[Takahashi et~al.(2026)Takahashi, Li, and
  Yoshida]{TakahashiLiYoshida2026}
S.~Takahashi, Z.~Li, and B.~Yoshida.
\newblock A no-go theorem for order-two {C}lifford electric-magnetic duality,
  2026.
\newblock URL \url{https://arxiv.org/abs/2610.02097v2}.
\newblock Version 2.

\bibitem[Yu et~al.(2017)Yu, Chen, Roy, and Teo]{YuChenRoyTeo2017}
X.~Yu, X.~Chen, A.~Roy, and J.~C.~Y. Teo.
\newblock Twofold twist defect chains at criticality.
\newblock \emph{Phys. Rev. B}, 96:\penalty0 205435, 2017.
\newblock \doi{10.1103/PhysRevB.96.205435}.

\bibitem[Ho et~al.(2015)Ho, Cincio, Moradi, Gaiotto, and
  Vidal]{HoCincioMoradiGaiottoVidal2015}
W.~W. Ho, L.~Cincio, H.~Moradi, D.~Gaiotto, and G.~Vidal.
\newblock Edge-entanglement spectrum correspondence in a nonchiral topological
  phase and {K}ramers--{W}annier duality.
\newblock \emph{Phys. Rev. B}, 91:\penalty0 125119, 2015.

\bibitem[Aasen et~al.(2016)Aasen, Mong, and Fendley]{AasenMongFendley2016}
D.~Aasen, R.~S.~K. Mong, and P.~Fendley.
\newblock Topological defects on the lattice: {I}. {T}he {I}sing model.
\newblock \emph{J. Phys. A}, 49:\penalty0 354001, 2016.
\newblock \doi{10.1088/1751-8113/49/35/354001}.

\bibitem[Pace et~al.(2025)Pace, Delfino, Lam, and
  Aksoy]{PaceDelfinoLamAksoy2025}
S.~D. Pace, G.~Delfino, H.~T. Lam, and {\"O}.~M. Aksoy.
\newblock Gauging modulated symmetries: {K}ramers--{W}annier dualities and
  non-invertible reflections.
\newblock \emph{SciPost Phys.}, 18:\penalty0 021, 2025.
\newblock \doi{10.21468/SciPostPhys.18.1.021}.

\bibitem[Howell(1986)]{Howell1986}
J.~A. Howell.
\newblock Spans in the module $(\mathbb{Z}_m)^s$.
\newblock \emph{Linear Multilinear Algebra}, 19\penalty0 (1):\penalty0 67--77,
  1986.
\newblock \doi{10.1080/03081088608817705}.

\end{thebibliography}
\end{document}